\documentclass[conference, 10pt]{IEEEtran}
\usepackage[utf8]{inputenc}
\usepackage{comment}
\usepackage{url}
\usepackage{verbatim}
\usepackage{enumitem}
\usepackage{multirow}
\usepackage{graphicx}
\usepackage{array}
\usepackage[linesnumbered,ruled,vlined,commentsnumbered]{algorithm2e}
\usepackage{setspace}
\usepackage{ifthen}
\usepackage{hyperref}
\usepackage{authblk}
\newcolumntype{P}[1]{>{\centering\arraybackslash}p{#1}}

\usepackage{microtype}
\everypar{\looseness=-1}
\usepackage{etoolbox}
\usepackage{tikzpagenodes}
\usepackage{pifont}
\DeclareRobustCommand*\circled[1]{\tikz[baseline=(char.base)]{ \node[shape=circle,draw,color=white,fill=black,inner sep=0.5pt] (char){#1};}}
\DeclareRobustCommand*\circledempty[1]{\tikz[baseline=(char.base)]{ \node[shape=circle,draw,color=black,fill=white,inner sep=0.5pt] (char){#1};}}

\usepackage{amsmath}
\newenvironment{nstabbing}
  {\setlength{\topsep}{0pt}%
   \setlength{\partopsep}{0pt}%
   \tabbing}
{\endtabbing}

\usepackage{listings}
\newcommand{\sysname}{\textsc{\footnotesize{IoTRepair}}}

\newcommand{\degree}{\ensuremath{^\circ}}
\def\ie{{i.e.},~}
\def\eg{{e.g.},~}

\newcommand{\para}[1]{\vspace{2pt}\noindent\textbf{#1.}}

\usepackage[flushleft]{threeparttable}
\usepackage{xcolor}

\usepackage[scaled=.8]{beramono}

\usepackage{microtype}
\usepackage{caption}

\usepackage{listings}
\usepackage[T1]{fontenc}
\usepackage[utf8]{inputenc}
\DeclareCaptionFont{black}{\color{black}}
\title{}
\title{\LARGE \textsc{IoTRepair}: Systematically Addressing Device Faults in Commodity IoT}
\author[1]{Michael Norris}
\author[2]{Berkay Celik}
\author[1]{Patrick McDaniel}
\author[1]{Gang Tan}
\author[1]{\\Prasanna Venkatesh}
\author[1]{Shulin Zhao}
\author[1]{Anand Sivasubramaniam}
\affil[1]{\small{Department of Computer Science and Engineering, Penn State University, State College, Pennsylvania} }
\affil[2]{\small{Department of Computer Science, Purdue University, West Lafayette, Indiana}}
\begin{document}
\newcounter{technicalDoc}
\setcounter{technicalDoc}{1}
\maketitle

\begin{abstract}
IoT devices are decentralized and deployed in unstable environments,
which causes them to be prone to various kinds of faults, such as
device failure and network disruption. Yet, current IoT platforms
require programmers to handle faults manually, a complex and
error-prone task.
In this paper, we present \sysname{}, a fault-handling system for IoT that (1)
integrates a fault identification module to track faulty devices, (2)
provides a library of fault-handling functions for effectively
handling different fault types, (3) provides a fault handler on top of
the library for autonomous IoT fault handling, with user and developer
configuration as input.
Through an evaluation in a simulated lab environment and with various
fault injection methods, \sysname{} is compared with current fault-handling solutions. The fault handler reduces the incorrect states on
average 50.01\%, which corresponds to less unsafe and insecure device
states.
Overall, through a systematic design of an IoT fault handler,
we provide users flexibility and convenience in handling complex IoT
fault handling, allowing safer IoT environments.
\end{abstract}


\section{Introduction}
The Internet of Things is constantly being developed to allow for a wide array of devices to be deployed and interpret, react to, and modify elements of a diverse set of environments. These devices are distributed in an environment, and the connectivity between them enables applications to manage some or all of the decentralized devices autonomously. This autonomous functionality can be used to provide services to users in home, industrial, and vehicular deployments. However, decentralized autonomous devices that interact with physical environments must maintain high dependability for practical deployment.

IoT devices are abnormally prone to diverse faults due to constraints such as minimal computational and energy resources, architectural problems, and disruptive environmental conditions~\cite{hnat2011hitchhiker,sharma2010sensor,szewczyk2004lessons, ni2009sensor}.  Faults can be caused by a complication in the device, such as loss of power or a bug in the software. A faulty device could become unresponsive or manifest an incorrect state. These faults lead to improper behavior with severe consequences, such as leaving a door unlocked,  allowing an adversary to break in the house, or leaving a valve open,  flooding a factory. Fault tolerance that targets individual devices and applications is insufficient, as interactions between applications can cause complex failures. 

The unique requirements for fault tolerance in IoT have been discussed~\cite{terry2016toward}, and prior works have proposed solutions  for fault identification~\cite{choi2018detecting,fang2013unifying, munir2014failuresense, sharma2010sensor} and addressing faults~\cite{ardekani2017rivulet,choubey2015power,kapitanova2012being,kodeswaran2016idea,su2014decentralized,tu2018redundancy, sengupta2019transactuations}. However,  previous solutions target a narrow scope, \ie only certain devices,  faults, or environments, and often only notifies users about the fault---assuming the user is familiar enough with the deployed devices,  installed applications, and fault types to act appropriately. Additionally, the solutions that do perform automatic handling rely on replication, which is not always effective \cite{tu2018redundancy} and may be prohibitively costly \cite{terry2016toward}. Since IoT systems are primarily autonomous, there is less interaction and oversight from users, which could lengthen the time to resolve the fault manually. The user must recognize the alert and respond to it, and there may not be a user on site. Therefore, an automatic response to faults is desirable to reduce dependency on swift user interventions and increase the reliability of IoT services. 



In this paper, we develop a flexible multi-layer fault-handling system called \sysname{} specifically designed for IoT. At the lower layer, \sysname{} provides a  fault-handling library with a set of functions for handling device faults and a configuration file. To handle diverse faults in IoT, users can write their own configuration files to customize the system and combine the fault-handling functions through the library's API. At the higher layer, \sysname{} includes an automated fault handler on top of the lower layer to handle common situations of IoT faults. The automated handler can be installed onto the edge device of a deployed IoT system to provide autonomous fault handling. A configuration file is generated at installation time through querying the edge device for a list of devices and applications, and also modified at runtime by the fault handler for runtime adaptation; e.g., the fault handler discovers redundant devices at runtime and saves that information in the configuration file.


To evaluate \sysname{}, we implemented 11 distinct IoT apps to manage 17 IoT devices in a simulated smart home.
We conducted three sets of experiments. The first is to measure the latency of different handling methods. The 
second is to measure effectiveness at reducing the errors in the system caused by injected faults. The third is to measure the energy overhead of the fault handler on devices. We inserted a comprehensive set of faults into 
the devices to measure errors that faults would cause that impact safety and security of the system, and how 
well the fault handler mitigated these errors. We evaluated the effectiveness in terms of incorrect states and 
energy overhead when a single fault and multiple faults were present. \sysname{} could reduce system errors by 
an average of 50.01\%. In this paper, we make the following contributions:

\begin{itemize}
\item We study the nature of faults in IoT systems. We show what makes fault behavior in IoT systems different from other environments and why addressing IoT faults is uniquely difficult and requires a generalized solution.
    
\item We design and implement a fault-handling library that implements a common set of functions for handling 
device faults, such as device restarts, retries, and checkpointing. We provide flexible APIs for developers to 
utilize our fault-handling library in their application code. On top of the fault-handling library, we develop 
an automated fault handler for IoT. The automated fault handler invokes fault-handling functions in customized 
orders and configures the functions automatically based on the IoT environment.
    
\item In designing \sysname{}, we propose a set of novel techniques, including a history-based 
checkpoint/rollback mechanism, and a technique for inferring redundant devices according to runtime 
information.

\item We evaluate \sysname{} on a simulated smarthome, including  17 devices and 11 IoT apps. We show what 
harm faults can cause and how \sysname{} mitigates the damage.
\end{itemize}


\begin{table*}[th!]
\def\arraystretch{1}
\setlength{\tabcolsep}{10pt}
\resizebox{\textwidth}{!}{%
\begin{threeparttable}[b]
\begin{tabular}{|l | l|l|l|}
\hline
\multicolumn{2}{|c|}{\textbf{Fault Type}}  & \textbf{Description} \\ \hline \hline
\multirow{3}{*}{\textbf{Fail stop}} 
& Power  & Device loses power from battery or outlet and ceases to function \\ \cline{2-3}
& Communication  & Device disconnects from network or otherwise cannot send or receive packets and ceases to respond requests  \\ \cline{2-3}
& Critical Error  & Hardware or software error causes device to cease to function \\ \hline
\multirow{4}{*}{\textbf{Non fail-stop}} 
& Outlier  & Device reports incorrect state for a single poll \\ \cline{2-3}
& Stuck-at  & Device fails to change state when expected to \\ \cline{2-3}
& High-Variance  & Device oscillates between states faster than environment dictates \\ \cline{2-3}
& Spike  & Numeric device state increases/decreases faster than environment dictates \\ \hline
\textbf{Other} & Cascading & Device enters incorrect state due to interaction with a faulty device through application code \\ \hline

\end{tabular}
\end{threeparttable}
}
\caption{Categorization of IoT device fault types and causes.}
\label{table:faultTypes}
\end{table*}

\section{Problem Statement and Motivation}
We begin by introducing the fault types and the reasons that IoT diverges 
from other computing platforms by focusing on the IoT platform architectures,
and then presenting an example of IoT implementation to illustrate the fault
types (Section~\ref{sec:faultsTypes}). Lastly, we study eight IoT programming 
platforms to understand their capabilities for fault handling 
(Section~\ref{sec:IoTSystemStudy}). Our study aims to characterize the current
state of fault tolerance in IoT systems and demonstrate the unique challenges
and design flaws that guide \sysname{}'s design.

\subsection{Faults in IoT Systems}
\label{sec:faultsTypes}
IoT systems integrate physical processes with digital connectivity. These
systems perform simple tasks such as motion-activated light 
switches, as well as complex tasks such as controlling traffic lights in 
a smart city. Regardless of purpose and complexity, IoT systems often 
use an edge device as a centralized gateway that connects devices in
physical environments and use a cloud back-end to synchronize device states 
and provide interfaces to control and monitor devices. 

Faults in IoT systems can occur due to flaws in components of devices, edge
device, cloud, and communication between them. While faults may happen in
any of these components, device failures are more common due to factors 
such as minimal computational resources, energy constraints, architectural
problems, improperly configured systems, and disruptive environmental 
conditions~\cite{padmavathi2009survey, kavitha2010security, sharma2010sensor, kapitanova2012being, ni2009sensor}. For instance, devices in smart homes could
experience faults on average two hours a day due to the power loss, network 
disruption, and hardware failure~\cite{ hnat2011hitchhiker}. In other veins,
experiments in exposed environments recorded half of the devices 
reporting incorrect states due to severe weather conditions~\cite{szewczyk2004lessons}.

\para{Scope} While fault tolerance includes techniques for both fault identification and fault handling~\cite{tanenbaum2007distributed}, in this work, we focus entirely on the fault-handling aspect and assume there is already a fault-identification module that can accurately identify the faults in a timely manner. The fault can be simple fail-stop faults, Byzantine faults, or faults injected by an adversary who has gained remote or physical access to a device.  At runtime, the fault identification module sends a signal with the faulty device ID to the handler when it detects a fault. The fault handler then determines the way the fault should be handled.

\begin{figure}[t!]
\centering
\includegraphics[width=1\columnwidth]{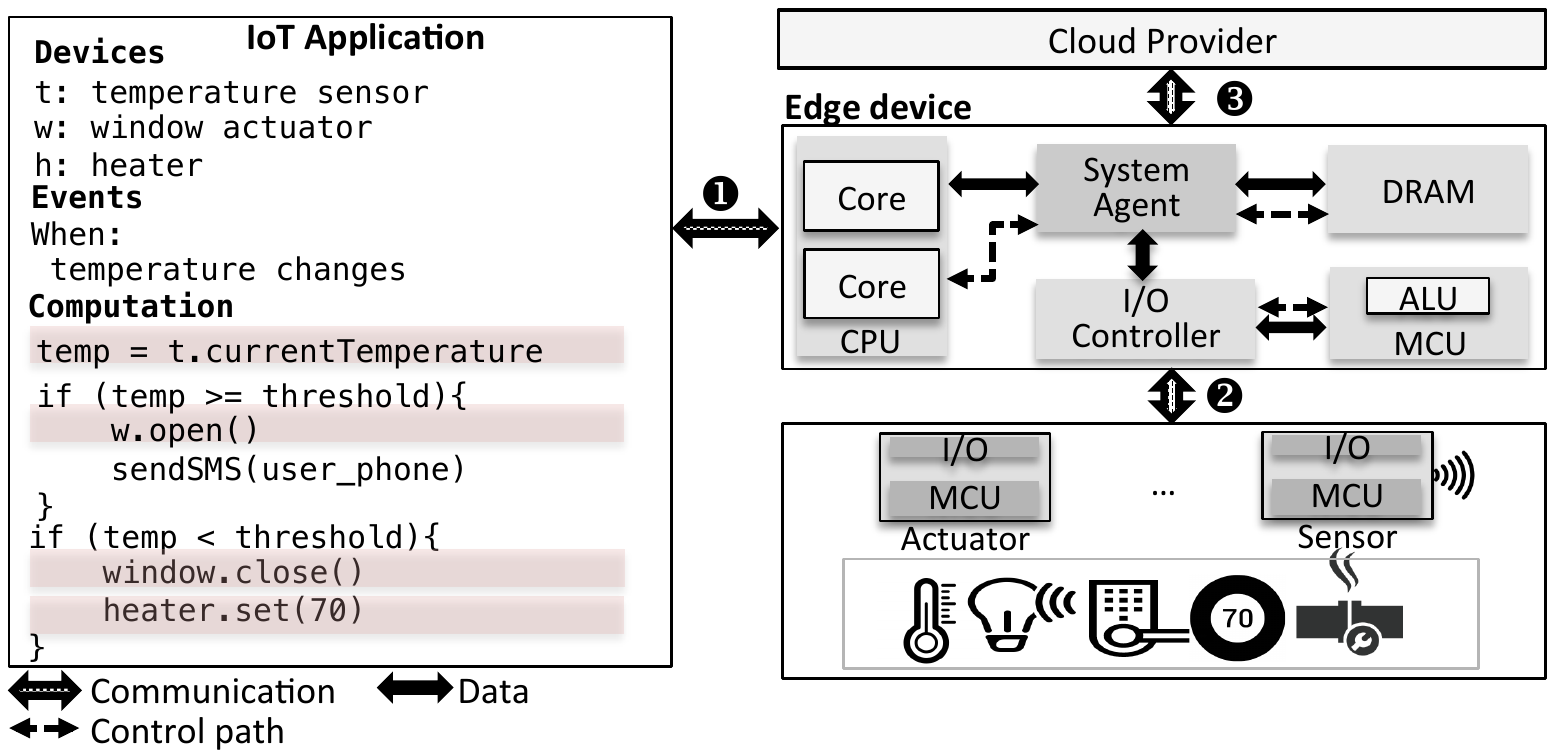}
\caption{IoT system architecture and illustration of fault types.}
\label{fig:motivation}
\end{figure} 

\para{Classifying Faults} An IoT device often consists of five
components (Figure~\ref{fig:motivation}, right): (i) a set of sensors such as location, temperature, and light 
sensors, (ii) an auxiliary Micro-Controller Unit (MCU) to read the raw sensor values, (iii) low
power CPU cores, (iv) network interfaces to communicate the user-level events to end-users, and (v) a battery 
or a power supply unit to power all these components. Components (ii) to (v) are often housed in an edge device such as a hub. A fault in any of these components can lead to errors in the system. We will use a 
\texttt{temp-control} app that subscribes to a thermostat, a heater, and window actuators in order to 
illustrate the different fault types and their consequences (See Figure~\ref{fig:motivation}). The app opens 
the windows and notifies the user through SMS when the temperature exceeds a user-defined threshold, and closes the windows and sets heater to a user-defined temperature value when the temperature is below the
threshold.


We divide faults into three categories extending the terminology in previous fault tolerance work~\cite{choi2018detecting} (See Table~\ref{table:faultTypes}). \emph{Fail-stop} faults happen when a device stops functioning and is unresponsive to external requests; for example, when a device loses power, network communication fails, or a software or hardware error halts device operation. A fail-stop fault in the thermostat would cause an app to halt entirely, and such a fault in the window or heater would remove the functionality of those actuators. 

\emph{Non-fail-stop} faults relate to a response by the device that diverges from the desired device state. 
These faults can manifest in a variety of ways with different effects, as shown in \cite{ni2009sensor}: an 
\emph{outlier} fault appears when the temperature falls outside of a reasonable range for a single device poll. An unhandled outlier fault in thermostat could cause the app to send an unintended notification that 
incorrectly indicates the temperature is above the threshold. A \emph{stuck-at} fault happens when a device 
cannot change state, maintaining the same state despite changes in environment or actuation commands. An 
example is if the application opens the window, but the temperature sensor fails to decrease the temperature as it should, which causes a safety issue because an open window enables a burglar to break in the house. A 
stuck-at fault could also cause energy data analytics to be incorrect and home temperature to surge if the 
heater gets stuck on. A \emph{high variance} fault appears when a device's state fluctuates back and forth more rapidly than the environment dictates. For example, the temperature value fluctuates between high and low more 
than the environment temperature. If this variance crosses the user-defined temperature threshold, this could 
cause several issues by repeatedly opening then closing the window and turning the heater on and off. A 
\emph{spike} fault happens when there are a rapid increase and decrease in temperature faster than the actual 
temperature in the environment. This could cause a safety issue by incorrectly opening the window for periods 
and turning on the heater and overheating the house---wasting energy and causing discomfort.


A faulty device may cause cross-application \emph{cascading faults}. Cascading faults happen when a faulty device in an app incorrectly triggers an event in another app. To illustrate, we consider a \texttt{secure-home} app co-located with the \texttt{temp-control} app. The \texttt{secure-home} app controls a presence sensor, a door lock, and a window actuator. The app keeps home secure---door-locked and window-closed---when the user is at not home and notifies the user if the house becomes insecure. An unhandled spike fault in a thermostat in the \texttt{temp-control} app causes an increase in temperature when the user is not home and causes the window to open. The \texttt{secure-home} app then sends an alert SMS to the user saying that the house is in an unsafe state---which makes the user become unnecessarily panicked and anxious. This behavior is correct from the \texttt{secure-home} app, as the window should not be open; however, the window is opened because of a fault that triggers the event handler of the \texttt{temp-control} app---faults that influence the physical spaces by a faulty device can propagate beyond the device through interactions through applications.


\begin{table}[t!]
\def\arraystretch{1}
\resizebox{\columnwidth}{!}{
\begin{threeparttable}[b]
\begin{tabular}{|c|c|c|c|c|c|c|c|}
\hline
& \multicolumn{3}{|c|}{\textbf{Fail-Stop Faults}} & \multicolumn{4}{|c|}{\textbf{Non-Fail-Stop Faults}} \\ \hline \hline
\textbf{IoT Platform} & \textbf{Comm.} & \textbf{Power} & \textbf{Crit Error} & \textbf{Outlier} & \textbf{Stuck-at} & \textbf{High Var.} & \textbf{Spike}\\ \hline 
\textbf{SmartThings~\cite{samsung}} & $\oslash$  & $\oslash$  & $\oslash$  & $\otimes$  & $\otimes$  & $\otimes$   & $\otimes$                   \\ \hline
\textbf{OpenHab~\cite{openhab}} & $\oslash$ & $\oslash$ & $\oslash$ & $\otimes$ & $\otimes$ & $\otimes$ & $\otimes$                   \\ \hline
\textbf{Vera~\cite{vera}} & $\oslash$ & $\oslash$ & $\oslash$ & $\otimes$ & $\otimes$ & $\otimes$ & $\otimes$                   \\ \hline
\textbf{Homekit~\cite{homekit}} & $\ominus$ & $\ominus$ & $\ominus$ & $\otimes$ & $\otimes$ & $\otimes$ & $\otimes$                   \\ \hline
\textbf{Wink~\cite{wink}} & $\ominus$ & $\ominus$ & $\ominus$ & $\otimes$ & $\otimes$ & $\otimes$ & $\otimes$                   \\ \hline
\textbf{AndroidThings~\cite{androidthings}} & $\oplus$ & $\oplus$ & $\oplus$ & $\otimes$ & $\otimes$ & $\otimes$ & $\otimes$                   \\ \hline
\textbf{IoTivity~\cite{iotivity}} & $\ominus$ & $\ominus$ & $\ominus$ & $\otimes$ & $\otimes$ & $\otimes$ & $\otimes$                   \\ \hline
\textbf{KaaIoT~\cite{kaaiot}} & $\ominus$ & $\ominus$ & $\ominus$ & $\otimes$ & $\otimes$ & $\otimes$ & $\otimes$                   \\ \hline

\end{tabular}
\begin{tablenotes}[para, small]
    \item [$\oslash$] Silent
    \item [$\ominus$] Generic Error
    \item [$\oplus$] Detailed Error
    \item [$\otimes$] Undetected
\end{tablenotes}
\end{threeparttable}
}
\caption{IoT platforms' response to different device faults. Undetected means that the fault is not recognized by the platform; silent means no message sent to applications; generic error means message does not contain fault information; detailed error means message contains fault information. }
\label{table:PlatformHandling}
\end{table}

\subsection{Fault Tolerance in IoT Platforms}
\label{sec:IoTSystemStudy}
We have studied eight major IoT platforms to characterize their fault identification and handling methods. The results for each platform are shown in Table~\ref{table:PlatformHandling}, with undetected meaning that the fault is not recognized by the platform, silent meaning that it does not inform the applications, general error meaning that applications are thrown an error that does not contain the fault type, and detailed error meaning that applications are given a thrown error that specifies the fault type. For instance, SmartThings provides developers APIs to obtain the current device state and can notify an app when the device state is changed~\cite{smartthings}, so that an app developer could check for the state changes before continuing execution. This requires additional developer effort, and if a developer neglects this check, an app would operate ignorant of whether a device failure exists. Additionally, if a failed device is queried, SmartThings returns the last known device state or some default value, which may potentially return stale sensor states when a device is unreachable. This enables an app to continue functioning, yet the stale states may inadvertently actuate incorrect device states.

The results in the table show that only Android Things gives enough information to handle fail-stop faults effectively, and no platforms give the means for applications to handle non-fail-stop faults. In contrast, \sysname{} is able to handle these types of faults automatically, and further provides a toolset to allow developers on any platform to perform their own handling.

\section{Approach Overview}
\label{sec:approachOverview}

\begin{figure}[t!]
\centering
\includegraphics[width=1\columnwidth]{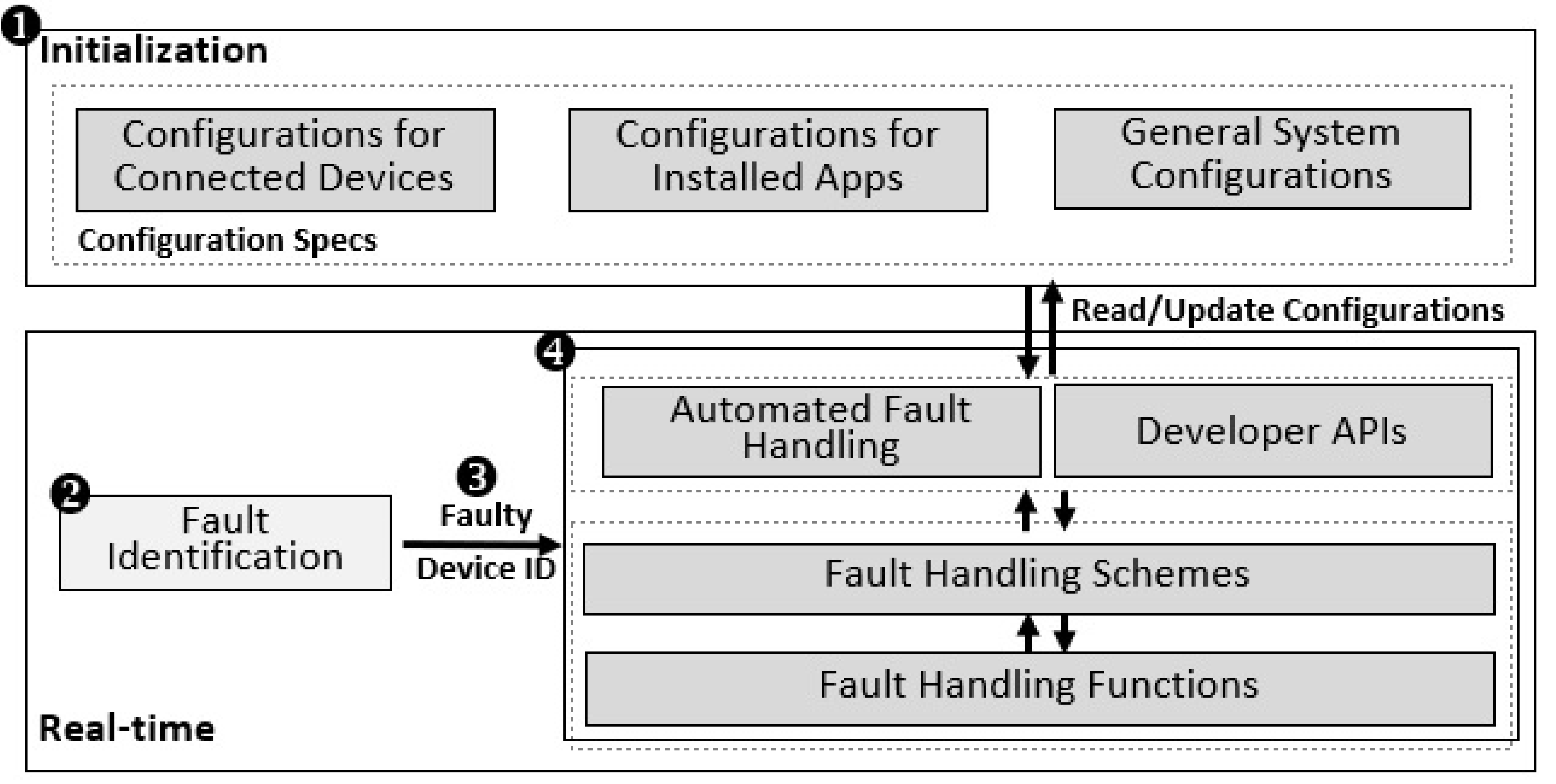}
\caption{Overview of \sysname{} architecture}
\label{fig:architecture}
\end{figure}

The variety of faults that can occur in IoT systems, combined with the diversity of IoT deployments, requires a flexible fault-handling approach. It is important that users can customize a fault-handling system to address a variety of faulty devices, fault types, deployment environments, and their preferences. For this reason, we design \sysname{} to be flexible, with major components presented in Figure~\ref{fig:architecture}.

First, the system provides a menu of fault-handling functions as a library so that developers can invoke these functions to mitigate or repair the faults. For instance, it provides a \emph{retry} function, which a developer can invoke to repeat some functionality of a device if a previous use of that functionality returned a fault. This is useful for resolving transient faults.






Second, the system provides an automated fault handler, which takes a configuration file (discussed soon) and performs fault handling without user involvement. It is parametrized by a fault identification
module, which detects faults and provides the ID of the faulty device (and optionally provides the fault type). In \sysname, we introduce the notion of fault-handling \emph{schemes} to organize and execute
fault-handling functions autonomously. A scheme is a list of fault-handling functions to be invoked in a specific order.  Each device is assigned a scheme in the configuration file that determines which fault-handling functions will be used to handle faults in that device, and what order they will be called in.

Finally, the system includes a {configuration file}, which is created during an initialization phase and updated by the automated handler at runtime. During the initialization phase, our system requests information about loaded identification module and obtains a list of installed applications and connected devices, their types, and capabilities; \eg a motion sensor has active and inactive states and hardware restart functionality. In detail, there are three types of information in the configuration file: \emph{(i)} general parameters, defining the upper bound on how long fault identification takes defined by the identification module, as well as specifications for checkpoint cleanup and replicated device detection, specified by the user. \emph{(ii)} device-based parameters, defining the list of devices running in the system and the parameters (e.g., which fault-handling scheme) for each device; a list of default configurations for various device types is defined in IoTRepair and each device has its default set from the matching type in this list, or to a conservative default if none exist default configurations are generated for each device connected to the edge device by selecting from a list of default configurations that best matches the device type; they can be modified by the user, by the automated handler at runtime, or by developers by making API calls in their applications; \emph{(iii)} application parameters, listing what apps are installed on the edge device and whether the application suppression is enabled for each; by default application suppression is disabled. These components allow \sysname{} to be flexible in addressing a variety of faults and environments. We next discuss two modes through which \sysname can be used.


\para{Developer API} In this mode, a developer can customize and use \sysname{} through the fault-handling library. The library provides an API for its functions (discussed later). An application developer can utilize the API in two ways. The simplest is to use a set of auxiliary functions to modify the configuration file to customize the automated handler. The second is for the application to call fault-handling functions in the library directly.  By using a try/catch block around application code that interacts with devices, the app can call handler functions in conjunction with their own code whenever a fault occurs in a device. This allows them to handle faults flexibly, although it is more effort on the developer part.



\para{Automated Fault Handling} \sysname{} can be loaded into an edge device. Once a faulty device is identified, the handler performs device suppression, which blocks polls to the faulty device and events generated by faulty devices. This prevents faulty devices from triggering application code and causing incorrect actuations. The handler then uses the configuration file to see which scheme should be used for handling the fault for this device. The handler then calls fault-handling functions in the order specified in the scheme, and each function uses the information in the configuration file while attempting to handle the fault. If the fault is repaired at any stage of the scheme, fault handling immediately ends, and the device is added back for normal execution. If the scheme ends without the device being repaired, the user is notified that the device must be manually repaired and the handling ends.

\section{Challenges in IoT Fault Handling}
\label{sec:challenges}
Compared to the fault handling in other computing platforms such as distributed systems and cloud services, IoT systems raise a set of unique challenges. We present these challenges below and how we address each.

First, there is no single fault handler function that can address all possible device faults, due to the vast array of heterogeneous IoT devices and the diverse set of environments in which they are deployed. Implementing an optimal handling technique for a specific fault is highly contextual---one cannot define the impact or correctness of fault handling without understanding the environment. For instance, different devices may require reordering functions to be executed to fit power or computation constraints, or a device might be more prone to either transient or permanent failures. To address this, we developed a set of functions that can be combined in any order depending on the device and configuration. We also created several built-in schemes that can execute these functions in different orders or exclude some entirely. These schemes can be chosen to meet the system and environmental requirements such as safety and security, and availability. We present the fault-handling functions and  schemes in Section~\ref{sec:functions}. 

The second challenge is device replication. In IoT deployments, environmental factors can make device functionality less reliable, and permanent failures in sensor readings and actuations are frequent. The goal of replication is to switch between primary devices and replicas without user intervention. To do this, we automate detecting redundant devices that are an exact duplicate through exploring sensor states on startup, and every time a new device is plugged in to the IoT system. This enables the system to fail-over from a faulty to the replicating device and to revert automatically when the device is no longer faulty. This also allows for multiple replications in case of cascading or simultaneous faults.  We discuss duplicate device detection and alternate devices to find devices that provide the same functionality in Section~\ref{sec:deviceBasedMechanisms}.

The last challenge relates to implementing practical rollback functions, which must consider not only the faulty device but all devices in an environment to determine a valid checkpoint for rollback. In general, rollback in distributed systems such as IoT is exceptionally challenging. Most rollback mechanisms ask each distributed process to checkpoint individually, and each process rollbacks to its recently used checkpoint~\cite{kshemkalyani2011distributed}.

In IoT environments, many factors in the system deployment determine the optimal rollback, such as the number of devices, correlation of device states, the urgency of fault correction, and consistency of system states. For instance, it is essential that the door is locked when the user is away, so a rollback would lock the door as soon as a fault was identified in the presence sensor. For this purpose, we implement a rollback scheme that rejects partial rollbacks; so if any device that requires a state change to match the checkpoint is faulty, the rollback is canceled. This method guarantees that the system is not in an undesired state no matter the implementation or fault. We will discuss our rollback in Section~\ref{sec:environmentBasedMechanisms}.

\para{Design Requirements and Assumptions} We present the requirements of \sysname{} to operate effectively. First, some fault-handling functions, such as rollback and checkpoint, require access to the device states; thus, we assume that device states are available through polling. Second, \sysname{} relies on the edge device (e.g., the hub) being able to query or record the ID and model of all connected devices. Third, we assume that there is a fault-identification module that provides the faulty device ID for each detected fault. Any system that can provide the faulty device ID can be integrated into \sysname{}. Finally, we assume that our implementation is injected into the edge device and it can access the aforementioned capabilities and its library functions are exposed to applications.

\section{Fault-Handling Functions}
\label{sec:functions}
We introduce a set of \emph{functions} to handle faults effectively. We present these functions in three groups, presented in Table~\ref{fig:MechanismPrototypes}. We then discuss four built-in fault-handling schemes that can be integrated into diverse IoT environments.

\begin{table}[t!]
\def\arraystretch{1}
\resizebox{\columnwidth}{!}{
\begin{threeparttable}[b]
\begin{tabular}{|c|}
\hline
\multicolumn{1}{|c|}{\textbf{Device-based Functions}}\\ \hline 
\texttt{bool activate\_redundant\_device (String device\_ID)\{\ldots\}}                        \\ \hline
\texttt{\begin{tabular}[b]{@{}c@{}}int retry (String device\_ID, FP verifyFunc\textdagger,\\String[][] expectedValues\textdagger, bool isFailstop\textdagger)\{\ldots\}\end{tabular}}                             \\ \hline
\texttt{bool device\_software\_restart (String device\_ID))\{\ldots\}}                     \\ \hline
\texttt{bool device\_hardware\_restart (String device\_ID)\{\ldots\}}                          \\ \hline
\texttt{None notifyUser (String device\_ID)\{\ldots\}}                              \\ \hline
\hline 
\multicolumn{1}{|c|}{\textbf{Environment-based Functions}}\\ \hline 

\texttt{bool checkpoint (String[] device\_values)\{\ldots\}}                                 \\ \hline
\texttt{int rollback (String device\_ID)\{\ldots\}}                                           \\ \hline
\texttt{int transaction (String[][] actuations)\{\ldots\}}                                       \\ \hline \hline
\multicolumn{1}{|c|}{\textbf{Auxiliary Functions}}\\ \hline 
\texttt{bool AddDevice (String[] device\_ID)\{\ldots\}}                                  \\ \hline
\texttt{bool RemoveDevice (String[] device\_ID)\{\ldots\}}                                    \\ \hline
\texttt{bool UpdateDeviceConfig (String[] device\_ID, ConfigOptions\textdaggerdbl)\{\ldots\}}                            \\ \hline
\texttt{bool UpdateAppConfig (ConfigOptions\textdaggerdbl)\{\ldots\}}                                     \\ \hline
\end{tabular}
\begin{tablenotes}[para, small]
    \item [\textdagger] Marks optional arguments. 
    \item [\textdaggerdbl] ConfigOptions represents a series of arguments for all configuration parameters for the device/application.
\end{tablenotes}
\end{threeparttable}
}
\caption{Fault handling functions prototypes.}
\label{fig:MechanismPrototypes}
\end{table}

\subsection{Device-based Functions}
\label{sec:deviceBasedMechanisms}
Device-based functions are an implementation of \emph{isolated} fault handling, which considers only the state of the faulty device, but not the overall state of the system.  The effectiveness of the function depends only on the specifications of the faulty device and the nature of the fault. The functions do not affect and are not affected by any other devices in the system.

\begin{figure}[t!]
\centering
\includegraphics[width=1\columnwidth]{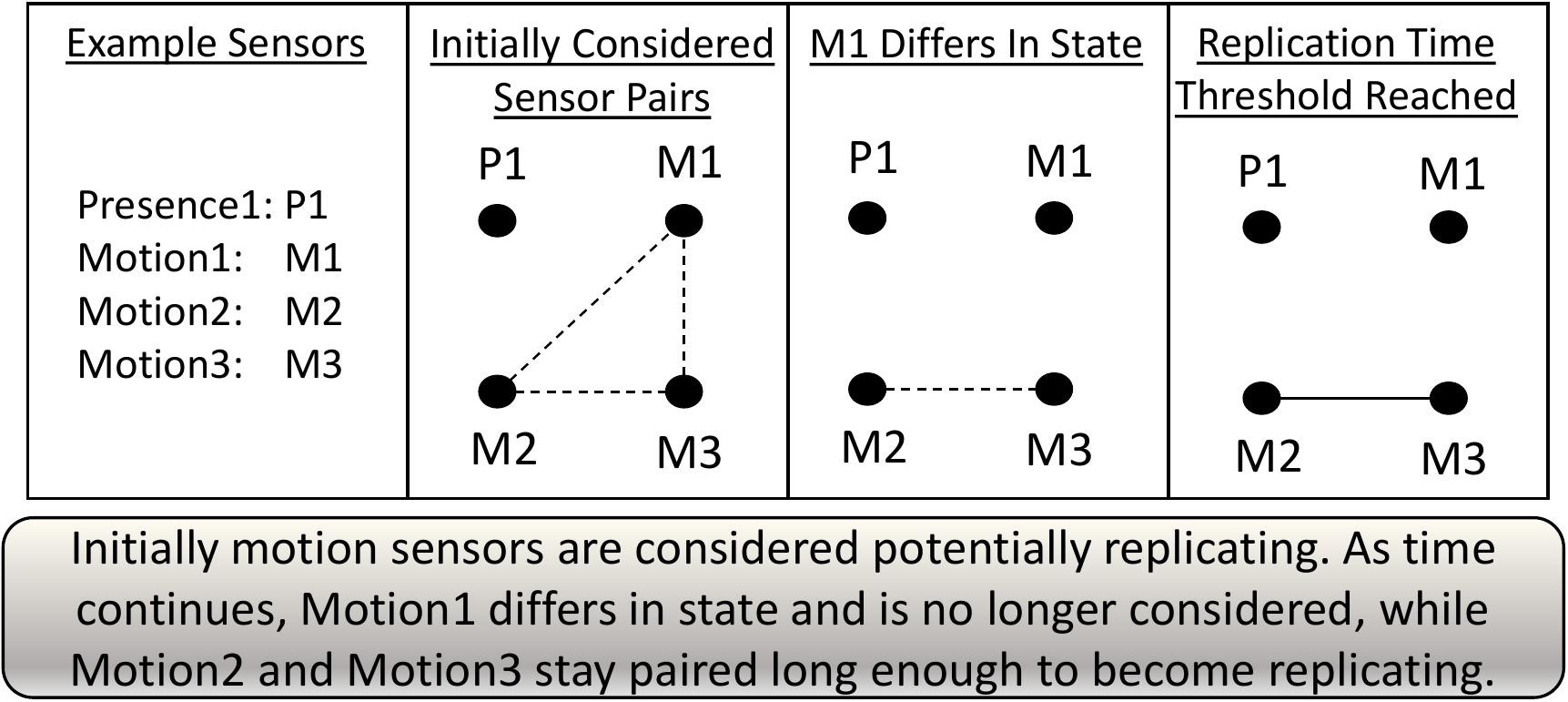}
\caption{Example of Automatic Replication Detection.}
\label{fig:replicationExample}
\end{figure} 

\para{Activate Redundant Device}  We define redundant devices to be of two types: (a) an exact replicate of the primary device;  for instance, a motion sensor may be a redundant device of another motion sensor; (b) providing the same capabilities as the primary device; for example, a security camera that is able to detect motion can be a redundant device for a motion sensor.

\sysname{} provides the support for both types by allowing the user to specify a device's replicate or same-capability devices in the configuration file.  The function \texttt{ActivateRedundantDevice} enables swapping active devices and allows for the system to continue to run unaffected by the detected fault, but requires the fault handler to know what devices can act as redundant devices for the faulty device. Specifically, it takes the faulty device ID as a parameter and then checks the configuration file to see if the device has redundant devices. If there is, it changes the device references in the edge device's connections so that polls and actuation commands to the faulty device are redirected to a replicating device.  Here, we use active replication~\cite{zheng2015selecting} to reduce the device activation overhead. In this, replicating devices are always active; thus, there is no additional computation to be performed after a device is replicated. In the case where more than one replicated device is set, the devices will be used in the order specified by the configuration.


To ease the burden of writing configuration files, \sysname{}'s detects redundant devices and automatically generates relevant configuration data. Currently, the support is only for type (a).  Finding a redundant device for type (a) in an IoT environment must consider the fact that similar devices are not necessarily co-located. To address this, we collect the sensor states and check whether devices report the same states in a given time threshold. If the two device states are the same and the transition probabilities converge with the events, these devices are identified to be replicating. This auto-generated duplicate device information can also be manually edited by the user for confirmation. Figure~\ref{fig:replicationExample} illustrates two co-located motion sensors detected as replicating as their states match consistently, while an unrelated presence sensor and distant motion sensor are discarded. To detect redundant devices of type (b), recent techniques such as using device fingerprints to automatically pair co-located devices~\cite{han2018you} could be easily integrated into \sysname{} as devices reacting to the context of their environment in similar manners means the fingerprint of one device can indicate what the value of the other device should be.


\para{Retry Device State} \texttt{Retry} is effective at handling short, transient faults in devices, stalling device execution long enough to let the fault resolve itself and preventing excessive handling from being performed. \texttt{Retry} takes parameters of the device ID and three optional arguments: a function pointer for a  validation function, a list of expected device states, and whether the fault is fail-stop or not. When used by the automated handler, these may be supplied by the fault identification module. Additionally, the configuration file specifies the max duration of retry for the faulty device. If no optional arguments are passed, the retry will delay other fault-handling functions for this length of time. If a function pointer is given, it is assumed that the function checks if the device is faulty, and it will be called continuously to determine if the fault is still present. If a list of expected values is passed, then the retry will poll the device continually for the duration; if the device returns the expected state several polls in a row, it will mark the fault as resolved and end fault handling. If the fault is identified as fail-stop, then any successfully returned state will be treated as the expected value. After the length of time expires, if the fault is non-fail-stop, the retry will disable the app and device suppression for the passed device and then wait for a time length equal to the upper bound on fault detection for the loaded fault identification module. If no fault is detected during this time, then the fault was transient and is marked as resolved, and fault handling ends. If a fault is detected or the fault was fail-stop, retry function terminates.


\para{Software and Hardware Restart} Restarting can be useful for either a software or hardware fault. \texttt{SoftwareRestart} and \texttt{HardwareRestart} take a device ID and send a signal to the  faulty device to initiate a software and hardware restart, respectively. The configuration file will also be read for the device to determine the max number of attempts to restart. If no acknowledgment signal is received, the command will be repeated, and if it is never acknowledged the restart will be aborted and return with an error code. If the restart is acknowledged, the function will continually poll the device to see when the restart is completed. Once the restart is completed, if the fault was not fail-stop, the function will disable the app and device suppression for the passed device and then wait for a length of time equal to the upper bound on fault detection for the loaded fault identification module. If no fault is detected during this time, then the restart resolved the fault and fault handling ends. If a fault is detected or the fault was fail-stop, the restart function ends unsuccessfully.

The implementation of restart functions are device-dependent. Many IoT market devices such as Honeywell Z-Wave  Thermostat~\cite{HoneywellThermostat} and open-source electronics platforms such as Arduino implement device restart functionality. 

\para{User Notification} \texttt{Notify} takes a device ID and notifies users of the fault and the result of fault handling. The implementation of the notify call is platform-dependent; we inform users through alert messages on the console, yet text messages or push notifications are possible through IoT platform APIs. Event triggers for user notification can be configured for each device to notify the user when a fault cannot be repaired by the handler when a fault has been repaired by the handler, when a fault occurs, or any combination of these.

\subsection{Environment-based Functions}
\label{sec:environmentBasedMechanisms}
Environment-based functions are an implementation \emph{linked fault handling}, which mitigates cascading faults. It responds to a fault by considering the states of all devices in an IoT environment, \eg linked fault handling should close the window that was incorrectly opened by an app due to the faulty thermostat. Our environment-based functions consider the state of all devices in the environment. As these functions can affect all device states in an environment, they are capable of mitigating cascading failures across the system.

\begin{figure}[t!]
\centering
\includegraphics[width=1\columnwidth]{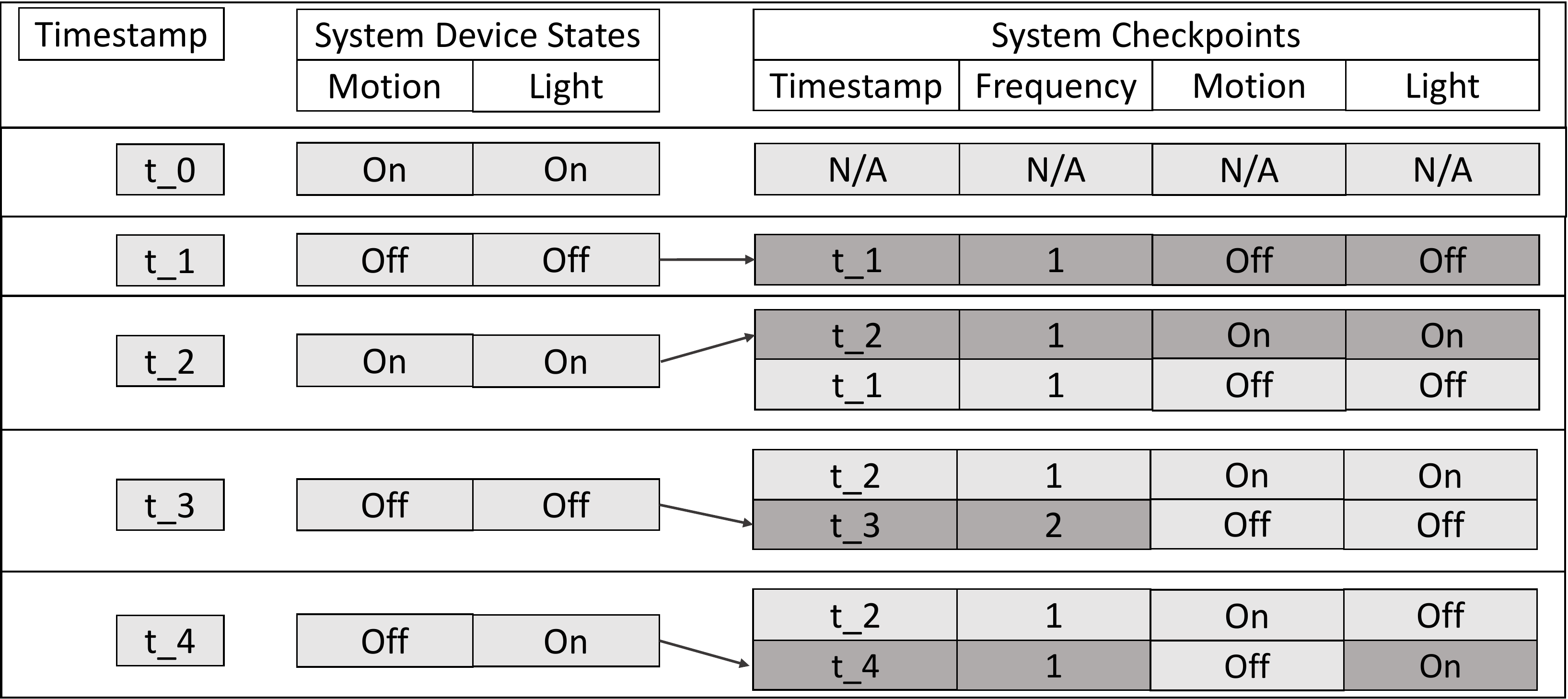}
\caption{Checkpoint illustration taken during system execution.}
\label{fig:checkExample}
\end{figure} 

\para{Checkpoint} \texttt{Checkpoint} stores all device states in the system at the time it is called.  We divide device states into \emph{sensor} states and \emph{actuator} states. Sensor states are read-only states that collectively represent the state of the environment and drive behavior. Actuator states can be read and modified through actuation, and they are the states that can be rolled back. 

We assume that there is a causal relation from sensor to actuator states; in other words, an app follows the well-known sensor-computation-actuator structure: it takes the sensor states, performs some computation, and then performs actuation to modify the actuator states. Based on this, \sysname{} includes a novel, history-based checkpoint/rollback mechanism (1) that during checkpointing records the history of device states, and (2) that during rolling back restores the most likely actuator states by looking up the history according to the current sensor states. We next discuss this mechanism in detail.

First, the automated fault handler invokes the checkpoint function after every actuation that does not trigger cascading actuations. This means a checkpoint is taken after actuations where no subscribing application initiates an actuation based on the new state. The handler determines these states by checking all apps logic subscribed to the actuated device to ensure no additional actuation is performed as a result. 

Second, after a checkpoint is taken, some time must pass before it is considered valid and appended to a history log of checkpoints. The delay period is determined by the upper bound on the fault-detection time so that the module can ensure that no faults were present at the time the checkpoint was taken.

When the delay expires, the new checkpoint is then stored in a history log, which holds checkpoints and their frequencies.  If no previous checkpoint has matching sensor states, a new checkpoint is appended to the log. If the sensor states match an existing checkpoint's sensor states, the frequency is incremented if actuator states also match, or the actuator states are overwritten, and the frequency is reset to be one if the actuator states differ. Only storing the most recent sensor states for a set of sensor states keeps the log from exploding in size. For the same reasons, checkpoints are removed from history if they remain unused for an extended time, as determined by the configuration. Figure~\ref{fig:checkExample} provides an example of how checkpoints are taken over time in a system as a result of changes in actuator states. The example uses a simple system with two devices: a motion sensor and a light actuator. The figure shows checkpoints being taken during a series of actuations, which occur at time-stamps $\mathtt{t_1}$ through $\mathtt{t_4}$, with initial time $\mathtt{t_0}$ before actuations. On the left side of the figure, the state of the system after the actuation completes is shown. On the right is the list of checkpoints at each time, starting from an empty set and updating after each actuation. The first two checkpoints at times $\mathtt{t_1}$ and $\mathtt{t_2}$ are new states, since there are no existing checkpoints that match the sensor states. For these new checkpoints, the time they occurred and current device states are recorded, and the frequency is set to 1. The checkpoint taken at $\mathtt{t_3}$ has the same device states as an existing checkpoint; so it updates the checkpoint time-stamp and the frequency. At time $\mathtt{t_4}$, a checkpoint that matches the sensor states is taken, but the actuator states do not match those of an existing checkpoint; so the matching checkpoint's time-stamp is updated, actuator states are overwritten, and frequency is reset.


\begin{algorithm}[t!]
    \DontPrintSemicolon
    \footnotesize
    \setstretch{0.9}
	\SetKwInOut{Input}{Input}
    \SetKwInOut{Output}{Output}
	\DontPrintSemicolon
	\SetNoFillComment 
	\Input{The $ID$ of the device that triggered this rollback}
	\Output{Success or Failure}
	\normalfont{Read configuration file to get }$\mathit{rollbackType}$ \;
	\normalfont{Get }$\mathit{targetCriteria}$ \normalfont{for }$\mathit{rollbackType}$
    $bestMatch \gets \emptyset$
    \BlankLine
    \normalfont{Search through} $\mathit{Checkpoints}$ \normalfont{to find checkpoint that fits} $\mathit{targetCriteria}$ \normalfont{and store it in} $\mathit{bestMatch}$ \;
    \If{$\mathit{bestMatch}$ \normalfont{equals} $\emptyset$}{$\mbox{return}$ $\mbox{Failure}$}
    \If{\normalfont{any }$\mathit{device}$ \normalfont{in }$\mathit{systemState}$ \normalfont{is an actuator, its state does not match} $\mathit{bestMatch}$, \normalfont{and is faulty}}{ $\mbox{return}$ $\mbox{Failure}$}
    \normalfont{Actuate each actuator in} $\mathit{systemState}$ \normalfont{to match} $\mathit{bestMatch}$ \;
    \normalfont{Change each faulty sensor state in} $\mathit{systemState}$ \normalfont{to match} $\mathit{bestMatch}$ \;
    $\mbox{return}$ $\mbox{Success}$
    \caption{Algorithm For Rollback\label{alg:Rollback}}
\end{algorithm}

\para{Rollback} \texttt{Rollback} performs a series of actuations to match the system state to the checkpoint that best reflects the current system sensor values. Algorithm~\ref{alg:Rollback} details the steps for computing \emph{rollback} that finds all potential matches, determines the best match and performs actuation to match the best match. 

We have designed three techniques for choosing the best checkpoint to target for rollback. The configuration for the faulty device determines which technique will be used. \emph{Most Recent} targets the checkpoint that has occurred in the system most recently. This technique is suitable for a system where faults can be detected quickly or system state changes slowly. \emph{Fail-safe} uses the fail-safe configurations for all of the actuators in the system to reduce the list of checkpoints to only those where actuators match their fail-safe states; e.g., the fail-safe configuration for an alarm might be to have it on. From this reduced list, a checkpoint whose sensor states match the current sensor states in the system is selected.  As the states of faulty sensors cannot be trusted, their sensor states cannot be considered when determining if a checkpoint matches the current sensor states. Because of this, it is possible that multiple checkpoints match the current known state of non-faulty sensors. If there is more than one match (or no match is found), then the frequency is used as a tiebreaker. \emph{Fail-norm} finds checkpoint matches in the same fashion as fail-safe, but does not reduce the list based on fail-safe configurations first and will not roll back to a checkpoint if none matches.

Rollbacks can be dangerous, as a partial rollback can result in an
otherwise impossible transition and enter an invalid system state. For this reason, our rollback fails if any actuator that needs to be changed is currently faulty (lines 6--7).  We note that this captures the heterogeneous nature of IoT devices when compared to similar functions in distributed systems~\cite{kshemkalyani2011distributed} and cyber-physical systems~\cite{kong2018cyber}.  We also implement a data rollback as part of the system rollback. Specifically, as long as the rollback does not fail, the values of any faulty sensors also set to their values at the time of the checkpoint until the device is repaired or another rollback occurs (Line 9). 

\begin{figure}[t!]
\centering
\includegraphics[width=1\columnwidth]{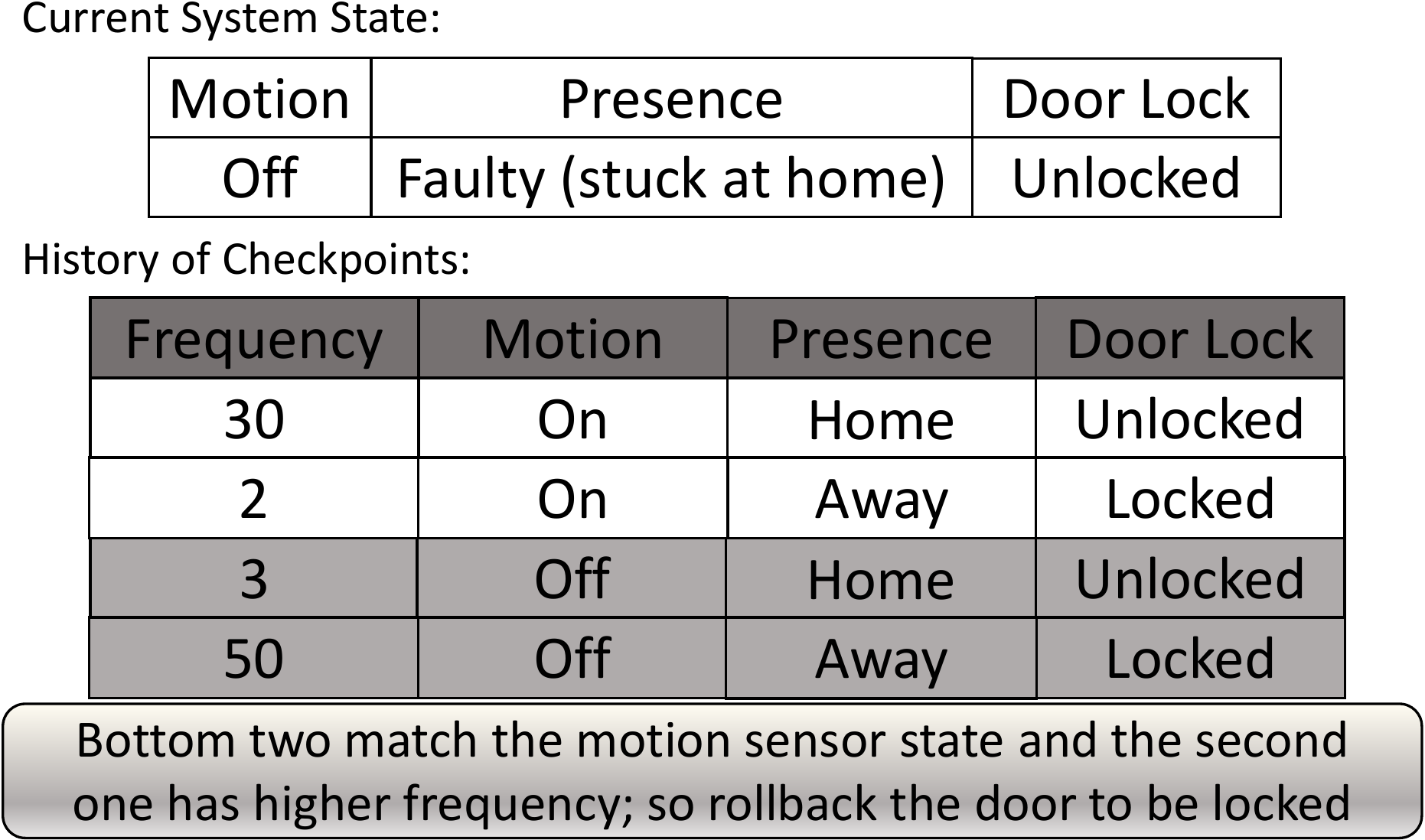}
\caption{New example of Fail-norm Rollback}
\label{fig:newrollbackExample}
\end{figure} 

Figure~\ref{fig:newrollbackExample} gives an example of how fail-norm
rollback would operate in a system with two sensors, motion and
presence, and a lock actuator for a door. It shows that \sysname{}'s
rollback can mitigate dangerous faults that would otherwise persist
until the user can repair the faulty device, as long as another
sensor's state is correlated with the faulty sensor's state with high
frequency. In the example of Figure~\ref{fig:newrollbackExample}, a
presence sensor that is stuck at \emph{home} could cause the door to
remain unlocked indefinitely. Fortunately, the motion sensor's state
is largely correlated with the presence sensor's state, because it is
likely to detect user motion when the user is at home. As a result,
\sysname{}'s rollback can then use the motion sensor's state to
correct the door to be locked and secure the home, even when the
presence sensor is faulty.


\para{Transaction} A transaction performs a series of actuations and ensures that the actuations are performed atomically as a group, either all successfully completing or none completing.  Currently, the \texttt{transaction} function is not automated, yet app developers can invoke it to cause a series of actuations to execute atomically when a partially executed series may leave the system in an unsafe state. For instance, a developer can issue a call ``\texttt{transaction([[Window, Closed], [Heater, On]])}'' to close the Windows and turn on the heater as a transaction. Transaction operates as a standard undo-log-based algorithm. It performs each of these actuations in the order given in the passed array while recording the original states in an undo log. If any actuation fails, the transaction is aborted, all devices are reset to their previous states, and an error is returned.

\subsection{Auxiliary Functions}
\label{sec:optionalMechanisms}
We describe functions necessary for the operation of the fault
handler. These functions are not directly used during fault handling but can be invoked by the automated handler or application developers to modify handler configuration. 

\para{Device Suppression} It is not a library function, but rather a
capability injected into the code of an edge device (e.g., hub). This is because it is required to interrupt polls and event hooks at the edge device. Through this capability, the edge device can terminate polling the device's
state and block actuation commands sent to the device. This prevents events from a faulty device triggering incorrect actuations to other devices through app code. Device suppression is effective in correcting non-fail-stop faults. For example, if the smoke detector experiences a high variance fault, it may rapidly switch between
smoke-detected and smoke-not-detected states. This would cause the alarm to turn on and off until the fault is resolved. With device suppression, the smoke detector can be suppressed once it is identified as faulty.

\para{Application Suppression} Similar to device suppression, application suppression is also a capability injected into the in the edge device code. Application suppression can be enabled or disabled for each app based on the configuration file. When a device is identified as faulty, application suppression checks the list of apps subscribed to the device's events. For every app that application suppression is enabled, all events that would be sent to the app and all commands generated by the app are suppressed. This halts the execution of the app, which is desirable when an app has functionality that may put the system into an unsafe state if the state of one of its dependent devices is unknown. For example, an app that opens the window when the presence sensor indicates the user is home and the house is too hot may open the window while the presence sensor is faulty if the temperature gets too hot.


\para{Update Configuration} This allows the automated fault handler to
update the configuration file at runtime. Configuration data for a
device or an application can be updated, if the passed arguments are
deemed valid for the configuration options.

\begin{table}[t!]
\def\arraystretch{1}
\setlength{\tabcolsep}{12pt}
\resizebox{\columnwidth}{!}{
\begin{tabular}{|c|c|c|c|c|c|c|}
\hline
\textbf{Scheme} &\multicolumn{6}{c|}{\textbf{Function Ordering}} \\ \hline \hline
\texttt{Conservative} & 1 & 2 & 3 & 4 & 5 & 6  \\ \hline
\texttt{Transient-resistant} & 1 & 3 & 4 & 5 & 6 & $\emptyset$  \\ \hline
\texttt{Long-Restart} & 1 & 2 & 5 & 3 & 4 & 6  \\ \hline
\texttt{Time-sensitive} & 1 & 5 & 2 & 3 & 4 & 6  \\ \hline
\end{tabular}
}
\caption{Execution order of functions in schemes. (1) Replicate; (2) Retry; (3) Software Restart; (4) Hardware Restart; (5) Rollback; (6) Notify User.}
\label{table:schemesOrders}
\end{table}

\subsection{Built-in Fault-Handling Schemes}
\label{sec:FaultHandlingSchemes}




\begin{table*}[th!]
\def\arraystretch{1}
\setlength{\tabcolsep}{1.1pt}
\resizebox{\textwidth}{!}
{%
\begin{threeparttable}[b]
\begin{tabular}{|l|l|l|c|}
\hline
\textbf{ID} & \textbf{App Name} & \textbf{Description} & \textbf{S/A\textdagger}  \\ \hline \hline
\texttt{App1} & Motion-Activated-Lights & When motion detected, turn on lights. Turn off lights when motion not active. & S.1, A.1  \\ \hline
\texttt{App2} & Smoke-Alarm & When smoke is detected sound alarm and unlock doors. When no smoke is detected turn off alarm. & S.5, A.2   \\ \hline
\texttt{App3} & Temperature-Control & Keep temperature between 70-80 degrees (\degree F) by turning heater and air conditioner on and off & S.3, A.2  \\ \hline
\texttt{App4} & Water-Leak-Detector & When leak is detected sound alarm and close water valve. When no leak is detected turn off alarm and open water valve. & S.6, A.2  \\ \hline
\texttt{App5} & Welcome-Home & When the user arrives home, unlock doors and turn on coffee machine & S.4, A.1 \\ \hline
\texttt{App6} & Secure-Patio & When user is not present and contact is detected, send text message to user & S.2, S.4, A.2  \\ \hline
\texttt{App7} & Energy-Saver & If window is open and either heater or air conditioner is on, close window. & A.1, A.2  \\ \hline
\texttt{App8} & Secure-Home & when user is not present home, lock doors and close windows. & S.4,A.1  \\ \hline
\texttt{App9} & Intruder-Detector & When user is not present home and motion is detected, send text message to user & S.1, S.4, A.2  \\ \hline
\texttt{App10} & Alarm-Safety & When alarm is activated, turn on lights & A.1, A.2  \\ \hline
\texttt{App11} & Morning-Air & Open windows and close windows at specific times  & A.1, A.2  \\ \hline
\end{tabular}
\begin{tablenotes}[para, small]
    \item [\textdagger] S is for Sensor and A is for Actuator device (Sensors and actuators are listed in Table~\ref{table:devices}.)
\end{tablenotes}
\end{threeparttable}
}
\caption{IoT apps, their sensors/actuators and descriptions, developed to evaluate \sysname{}.}
\label{table:applications}
\end{table*}

We introduce a set of built-in schemes to automate the functions above (See Table~\ref{table:schemesOrders}). These schemes modify the execution order of the functions to address requirements such as safety and security and real-time fault detection and correction. We chose to use the device type as a primary determinant of each scheme as IoT devices introduce several limitations, \eg duration of restarts, and whether a replicated device is available, which impacts the optimal functions ordering. We created four schemes for our evaluation based on properties we discovered across our device set. These are not designed to address all possible deployments, and additional schemes can easily be created considering the deployment requirements as new handling functions are developed. 

We  describe the purpose of each scheme and show the order they execute the functions. Each scheme starts with trying to activate a redundant device to see if the fault can be repaired. If not, the \emph{conservative} scheme then uses retry to fix the fault, which will be successful if the fault is transient. This fits in environments where there are no strict time requirements. The \emph{transient-resistant} scheme is similar to the conservative scheme, yet it skips the retry function. This aims at devices that are unlikely to experience transient faults. For instance, this would be suitable for a temperature sensor deployed in a stable home environment, since it is capable of restarting and does not control time-sensitive operations. The \emph{long-restart} scheme is used in devices that have excessively long software and hardware restart times, such as security cameras and a smart refrigerator. The restart functions are moved to the end of the scheme so that other functions are attempted prior to the long restart. The \emph{time-sensitive} scheme removes retry and moves rollback ahead of the restart functions. It aims to return the system to the desired state as the device impacts the safety, for instance, when the security system is unresponsive. This scheme fits industrial or vehicle environments where many devices are deployed, and system integrity is the top priority. This would be implemented in devices like smoke alarms, which could cause a fire hazard if down for even a short time.

\begin{table}[t!]
\def\arraystretch{1}
\setlength{\tabcolsep}{1.5pt}
\resizebox{\columnwidth}{!}{
\begin{threeparttable}[b]
\begin{tabular}{|l|l|c|c|c|c|}
\hline
\textbf{ID\textdagger}  & \textbf{Device\textdaggerdbl} & \textbf{Power (mW)} & \textbf{Read (ms)} & \textbf{Output} \\ \hline \hline
S.1  & Motion sensor & 66  & 0.1 & double, 8B \\ \hline
S.2  & Contact sensor & 0.1 & 0.1 & int, 4B \\ \hline
S.3  & Temperature, pressure, altitude sensor & 19.5 & 37.5 & double, 8B \\ \hline
S.4  & Presence sensor  & 1.3  & 0.5  & int*3,12B \\ \hline
S.5  & Smoke detector & 30 & 0.96 & double, 8B \\ \hline
S.6  & Leak detector & 80 &0.1 & double, 8B \\ \hline
A.1  & Door lock, coffee machine, light & 0.01 & 0.1 & - \\ \hline
A.2  & Alarm, air conditioner, heater, window, valve & 100 &0.1  & - \\ \hline
\end{tabular}
\begin{tablenotes}[para, small]
    \item [\textdagger] S is for Sensor and A is for Actuator.\newline
    \item [\textdaggerdbl] We use Arduiono compatible devices to simulate the devices in IoT apps. For instance, a flame detector is used for smoke-alarm, and switches are used for the lights (see Section~\ref{sec:implementation}).
\end{tablenotes}
\end{threeparttable}
}
\caption{Details of sensors/actuators used in our IoT apps.}
\label{table:devices}
\end{table}

\section{Implementation}
\label{sec:implementation}
We simulated a smart home in a lab setting using a set of IoT devices and applications. In this section, we begin by introducing our simulated devices and applications, and then present fault injection and fault identification techniques required for our evaluation of \sysname{}.

\para{Implementation Setup} We deployed a set of Arduino devices that were connected to and run in parallel on an AVR Arduino Mega 2560 Rev 3 Board~\cite{arduinoMega2560}. We utilized a mixture of analog and digital devices. The initial data collection was run until 50K device states gathered to collect the patterns of device behaviors. The polled data was output into a file that stores the behavior of sensors and actuators as the environment changes were observed. We use the sensor data to generate a realistic trace through simulated devices that represent a range of activity in a smart home. We use this trace of sensor states as input to a trace-driven simulation, which simulates the behavior of eleven IoT applications (detailed below).

\para{IoT Devices and Applications} We deployed six different types of sensors and two types of actuators, for a total of eight types of Arduino devices. Table~\ref{table:devices} presents the types of devices by ID, characteristics of each including the power consumption, read time for states, and the type of sensor output
values.  These devices are used in a simulated smart home environment where IoT applications are deployed. In the simulated environment, we deployed seven sensors, one for each type of sensors and also a replicated smoke detector. We also deployed 10 actuators of the two types, as our environment required a higher number of actuators than we had available in our lab. Table~\ref{table:applications} shows the description of
applications and devices used in each of them.  These apps are selected to cover all spectrum of home environment functionality, including green living, convenience, home automation, security and safety, and personal care.

\para{Fault Injection} In our evaluation, we use fault injection to evaluate the effectiveness of fault handling. There is an explosion of possible fault scenarios that can occur in an IoT environment due to
the many factors that can influence a fault.  Our fault injection aims to take into account these factors, including fault type, fault length, device type, whether a fault can be repaired automatically, the state of other devices when the fault occurs, and the events that occur during the fault's presence.  First, we inject faults evenly across all devices. Second, we specify that a fault can be repaired through software/hardware restarts, with equal chances.

The fault type, fault length, device states, and events are injected within reasonable distributions because these values impact the number of incorrect states that occurs from faults, rather than how the faults should be handled. After we have specified the fault injection parameters, we construct an input file that specifies when and what faults are injected and optionally when and what faults are removed in the trace. This allows our simulation to accept custom sets of faults to represent diverse IoT environments. Here, we generate \emph{permanent} and \emph{transient} faults based on the specifications of each evaluation (Section~\ref{sec:Evaluation}). Permanent faults are characterized through arrival time, length, type, and false value of every desired injected fault. For example, an injected fault entry, \texttt{(1000, 0, UNFIXABLE, 1)}, injects a fault at the 1000th poll time into the motion sensor (with device ID \texttt{1}), it cannot be repaired (UNFIXABLE), and it causes the devices to be stuck at value motion-detected (device state being 0).  For transient faults, an additional line must be added, such as \texttt{(2000, 0, NO\_FAULT, 0)}, which removes the fault from the 0th device at poll time 2000. Faults are injected or removed at the beginning of each 1-second polling cycle before any application or handler code is run.


\para{Fault Identification} As discussed earlier, \sysname{} is parameterized by a fault identification module. There are a wide array of identification algorithms available, discussed in Section~\ref{sec:relatedWork}. In our evaluation, we choose to use a perfect identification module that identifies faults instantly and correctly to show the maximum possible benefit of \sysname{}.

\section{Evaluation}
\label{sec:Evaluation}
We present our evaluation on the performance of \sysname{}'s fault-handling functions and automated handler (Section~\ref{sec:runningTime}). Furthermore, we assess the effectiveness of \sysname{} through injecting single and multiple faults (Section~\ref{sec:effectiveness}). Finally, we evaluated the power consumption of each run, and found that \sysname{} reduced power consumption over not having any handler. The detailed results of power consumption evaluation can be found in Appendix ~\ref{apdx:PowerCons}.


We discuss a couple of notes before we proceed. First, unless otherwise specified, we assume that the smoke detector and smart lights have replicated devices. These devices were chosen to demonstrate device replication because smart homes are likely to deploy multiple such devices. Second, in our evaluation, we found that the fail-norm checkpoint is consistently the best match for all devices. We also found that several devices, such as the contact sensor, did not have sufficient correlation with other devices for an effective rollback. Therefore, we used fail-norm or no rollback accordingly across our devices.

\begin{figure}[t!]
\centering
\includegraphics[width=\linewidth]{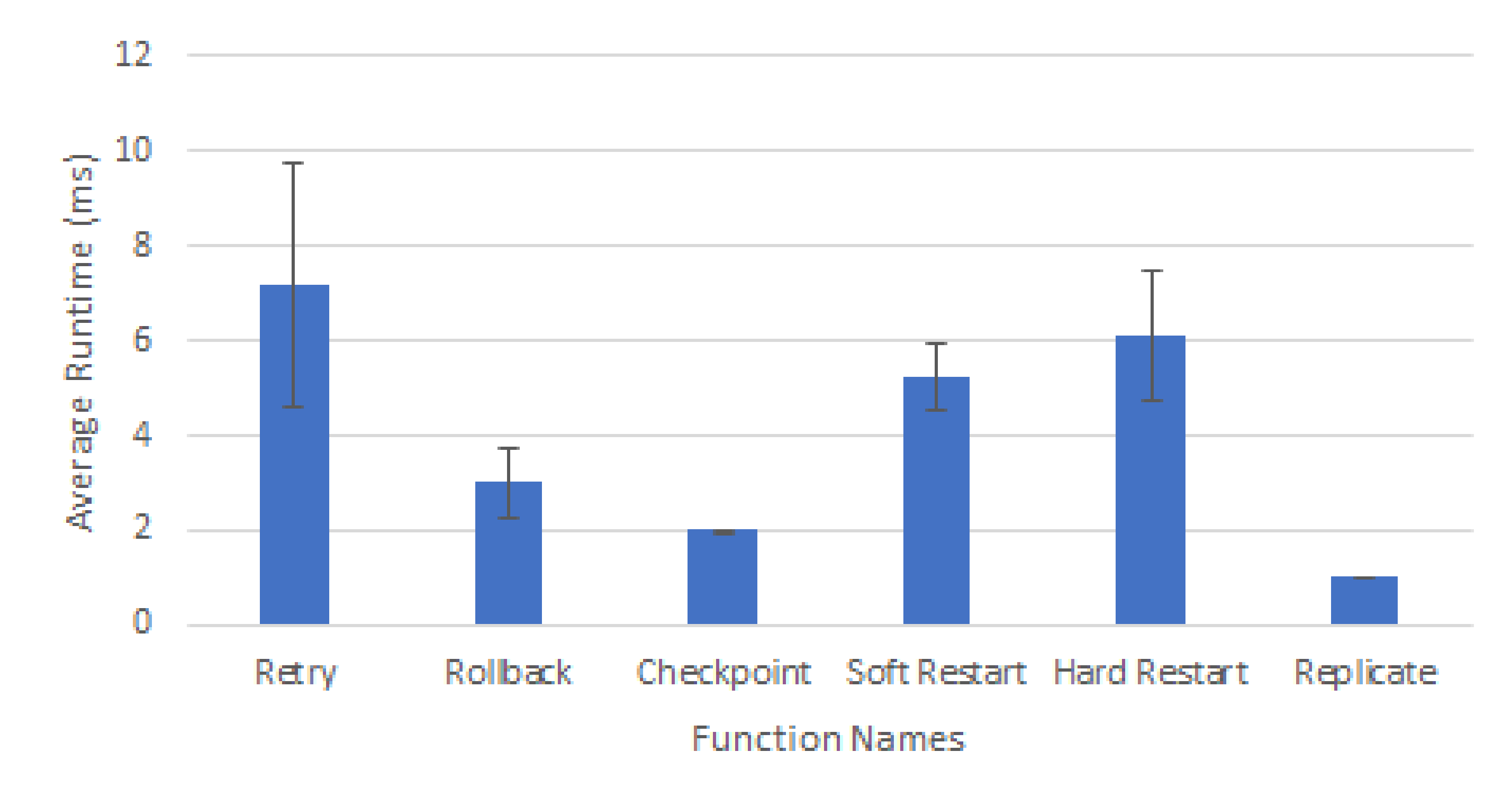}
\caption{The numbers illustrate the avg. time for each function to execute in msecs. This time is averaged across simulated devices, possible function parameters, and runtime environments.}
\label{fig:FunctionTiming}
\end{figure} 

\subsection{Fault Handling Latency}
\label{sec:runningTime}
In this experiment, we evaluate the latency of functions in \sysname{}'s fault-handling library and also the latency of its automated handler when using a specific fault-handling scheme. To get timing information, we examine the logic of a fault-handling function to track CPU operations and device interactions. Time spent on CPU operations is then calculated based on the processing speed of a representative SmartThings IoT Hub~\cite{smartHub}.  Time for device interactions is determined by device manuals for the simulated physical devices, or our implemented Arduino devices when datasheets are insufficient.




\para{Function Timing}
For this evaluation, we analyzed our fault-handling functions to get the CPU operations and device interactions performed across all devices, parameters, and fault types. We then calculated the running time of each execution, and the mean and standard deviation across all executions of each function. We used permanent faults and short (under 1 second) transient faults to represent the most likely fault conditions. We also did not consider fault-identification time when calculating runtime as our work focuses on fault handling.

The results are illustrated in Figure~\ref{fig:FunctionTiming}.  Retry takes the longest to execute because it always waits for either the duration of the fault, the device based timeout, or at least one read of the faulty device's state to confirm that the fault is resolved. In an environment with longer fault durations and longer timeouts, retry's runtime increases; for this reason, retry's runtime has a large standard deviation.  Conversely, checkpoint, and replicate's runtime are mostly stable; their operations are independent of devices and fault type and are primarily composed of CPU operations. Actuation commands dominate Rollback's runtime, and so the deviation is caused by how many devices require actuation to rollback to the selected checkpoint. In our example setup, the average number of actuations is 1.76; so the rollback time can increase for deployments with more devices.  Software and hardware restart vary depending on device restart time and how many restart commands are sent. They vary a little across devices but do not vary much across system deployments, as most devices have similar restart times. 

From these latencies, we find that all of our functions are fast enough to handle faults effectively. Reading a device state for most Arduino devices takes 6ms, and this is even longer in a real home deployment with distributed devices that must communicate over a network. This means most of our functions consistently take less time to execute than a device-state read. 


\begin{figure}[t!]
\centering
\includegraphics[width=\linewidth]{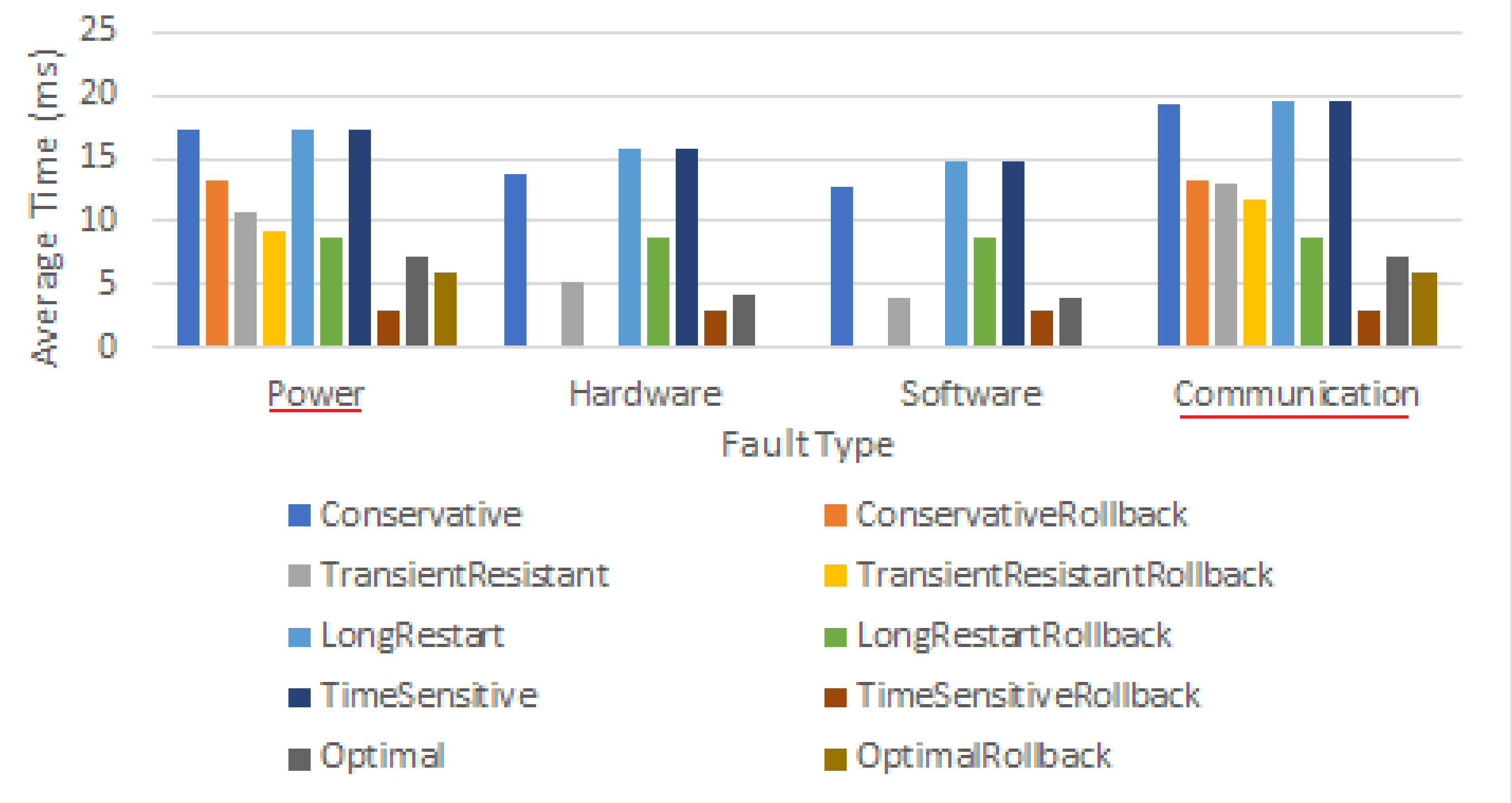}
\caption{Avg. timings in msecs to illustrate how long each scheme takes to handle different types of faults, and how long each scheme takes to rollback.}
\label{fig:TimeToHandle}
\end{figure} 

\para{Time to Handle Faults} 
We evaluated \sysname{}'s automated handler by organizing the fault-handling functions into the schemes defined in Section~\ref{sec:FaultHandlingSchemes}. We calculated the running time across all possible permutations of devices, function parameters, fault types, and schemes to obtain the average time each scheme takes to handle each fault type. We aim at evaluating the completion time of schemes for different fault types such as power and hardware. The results are shown in Figure~\ref{fig:TimeToHandle}, where the average time to handle a fault is shown for each scheme, alongside the average time each scheme takes to rollback. Rollback is the only form of fault mitigation that takes significant time. The rollback time to complete is dependent on where in the scheme rollback is called and system factors such as the checkpoint list size; therefore, we show rollback time for each scheme separately. In the graph, optimal is the time it will take if fault-handling functions were chosen with perfect knowledge of the fault and system. Using the proper scheme can result in repairs or mitigation occurring over twice as fast. In the figure, power and communication faults are underlined in red since our handling techniques cannot repair the fault. In these cases, rollback time is more critical, as mitigation techniques can still remove some errors caused by faults. For these faults, the handling time is only an indication of how long it takes the handler to recognize that the fault is unfixable and notify the user. We also found that, for most cases, it may take 10ms or longer to handle the fault, so \sysname{} is most useful for faults that last longer than a few milliseconds. However, even for short faults, suppression and replication help mitigate errors.



\subsection{Effectiveness}
\label{sec:effectiveness}
To evaluate the effectiveness of \sysname{}, we measure the number of \emph{incorrect devices states} that occur over executions. A device is in an incorrect state when it differs from the state it was in at that time in the faultless execution, for the same input. This metric is used to show the effectiveness of \sysname{} in reducing the effects of faults in devices. Note that this metric is a conservative definition since a fault handler may correct a device to a state acceptable to users even if it differs from the faultless execution state. However, we use the incorrect-state metric since it does not require an acceptable-state definition, which is user and application-specific.

For evaluation, we execute the simulation: \textcircled{a} when no faults are present, \textcircled{b} when faults occur without a fault-handling system, \textcircled{c} when faults are addressed with device suppression, and \textcircled{d} when \sysname{} handles faults. The baseline of the correct device states is obtained through the execution of \textcircled{a}. We evaluate the effectiveness of the four fault-handling schemes discussed in Sec.~\ref{sec:FaultHandlingSchemes}, and compare the baseline with each experiment by computing the number of incorrect device states caused by the injected faults across each of the executions. The reduction in incorrect states towards execution \textcircled{a} evaluates the effectiveness of each scheme's fault handling. We detail our findings when a single fault and multiple faults occur and \ifnum \value{technicalDoc}>0 {The Table ~\ref{table:experimentResults} in appendix ~\ref{apdx:SafetySec} gives examples of injected faults in the system and their effects in each execution.} \else {give a more comprehensive analysis of faults in the system and their effects in each execution in our extended paper~\cite{IoTRepairExtended}.}\fi




\begin{figure}[t!]
\centering
\includegraphics[width=\linewidth]{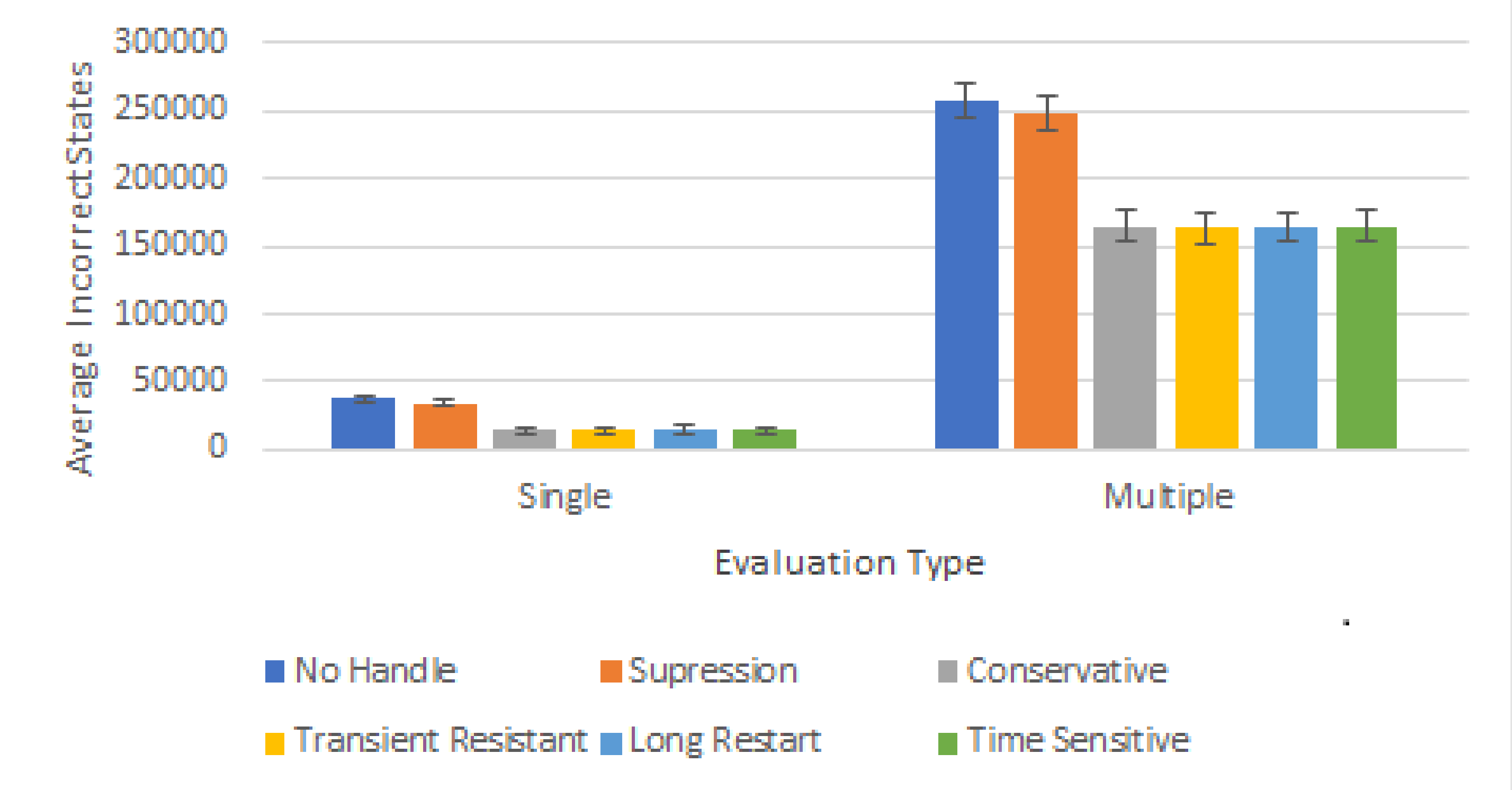}
\caption{The number of incorrect states in each execution type for Single and Multiple fault evaluations. The results show a significant decrease in incorrect states from No Handle and Suppression only to our fault handling schemes.}
\label{fig:incorrectStates}
\end{figure} 

\para{Single Faults} 
In our first effectiveness experiment, we consider single faults--when no more than one fault is in the system at any given time. This is typically caused in an IoT system when a single device runs out of battery, gets damaged by the environment, or experiences a communication interruption.  The ``Single'' section of Figure~\ref{fig:incorrectStates} presents the number of incorrect states over different executions. The transient-resistant scheme was the most effective scheme for handling single faults. It shows a 63.51\% decrease in incorrect states over no handler and a 60.43\% decrease over device suppression; this represents significant mitigation of the effects of faulty devices. Other schemes, although performing slightly worse than the transient-resistant scheme, also show a significant improvement over the cases of no handler and suppression.




We found during our experiments that our fault-handling functions can cause incorrect states that are not present in the no-handler case. 22.84\% of the incorrect states in the transient-resistant execution are caused by the handler and occur because of an incorrect rollback or because the system with no handler happens to be correct by coincidence. For the first reason, our rollback chooses a checkpoint most likely to be correct based on prior states and configuration, but this can cause incorrect states if the system was in an undesired state at the time the checkpoint  was taken, or if the usual type of rollback for this device is wrong for the current fault or environmental conditions. 
For the second reason, the faulty state of a sensor may happen to match what the correct state should be, but \sysname{} decides to suppress events from faulty devices. An example of this is an error where the faulty temperature value causes the heater on, but during the time it takes \sysname{} to fix the fault, the temperature falls, changing the environment so that the heater being on is correct. In this case, the no-handler execution turning on the heater achieved the correct state while \sysname{} causes an incorrect state.

To conclude, the fault-handling functions reduce the incorrect states over device suppression and when no fault handling is used. For the incorrect states that the handler does not cause, the state cannot be corrected through any means we are aware of. This is because no handling function can perfectly solve a core device experiencing a permanent fault, and faulty states can occur after fault is identified, but before fault handling completes. For the incorrect states caused by \sysname{}, our evaluation shows that suppression and rollback remove many more incorrect states than they generate.


\para{Multiple Faults} 
In our second set of experiments, we consider multiple faults---the case where there are two or more faulty devices in an IoT environment at a time. To do this, we inject additional faults into our single faults experiments to cause cases where possibly all devices can have fail-stop and non-fail stop faults simultaneously. We primarily inject fail-stop faults in this evaluation, because power outages and connection disruptions are common ways for multiple faults to occur at once. However, it is also possible for hazards in the environment to causes simultaneous non-fail stop faults, so we inject some non-fail stops as well in this evaluation. 

Multiple device faults increase the incorrect states, cause more unreliability in devices, and cause our handling functions to be less effective.  The ``Multiple`` section in Figure~\ref{fig:incorrectStates} shows the number of incorrect states caused by multiple faults. This scenario decreases the reduction in incorrect states from the no-handle case; the transient-resistant scheme reduces the number of incorrect states by 36.68\%.  It still performs the best, but it showed less than a 1.00\% improvement over the other schemes.

While multiple faults cause more incorrect states that can potentially be fixed, both our redundant device functions and rollback are less effective. Multiple faults can cause failures in redundant devices at the same time as the devices they are backing up. Also, checkpoints become less accurate for use in rollback, as faulty devices cannot be used when finding a matching checkpoint, thus increasing the number of possibly matching checkpoints to choose. Finally, as a system approaches total failure, partially successful handling may have a reduced effect or none at all. For example, multiple faults cause both the leak sensor and the alarm actuator to be faulty, affecting \texttt{Apps 1-4,6,7,10}.  If only one of the devices were faulty, handling the fault would correct the state of the alarm. With both devices being faulty, correcting only one device has no impact on the alarm's correctness.  Therefore, the more concurrent faults in the system, the less impact a fault handler will have. Even so, in this worst-case, \sysname{} still substantially reduces the effects of faults.

\para{Cascading Faults}
Our applications are susceptible to cascading faults---which are cases where a fault in one device causes incorrect states in a second device through inter-application interactions.  These cascading faults can be unpredictable because the application that places a device in a faulty state might not directly utilize the value of the faulty device, and the application interaction may not even be intended. For example, a temperature sensor becoming stuck at a low temperature causes the following sequence of events and actuations in \texttt{App3}, \texttt{App7}, and  \texttt{App11}:

\vspace{3pt}

{\footnotesize{
\begin{nstabbing}
\textbf{
heater-off$\xrightarrow{\text{temp<60}}$\texttt{App3}: heater-on $\xrightarrow{\text{heater-on}}$\texttt{App7}:close-window
}
\end{nstabbing}}}

\vspace{3pt} \noindent which contradicts another completely separate sequence:

\vspace{3pt}

{\footnotesize{
\begin{nstabbing}
\textbf{
window-closed$\xrightarrow{\text{time: 7am}}$\texttt{App11}: open-window
}
\end{nstabbing}}}
\vspace{3pt}

In this example, \texttt{App7} causes the window to close as a cascading result
of  \texttt{App3}'s action responding to a fault in the temperature sensor. This contradicts 
\texttt{App11}, which dictates that the window should be open due to the current 
time. Because of this, even if the temperature sensor is repaired, the system may
not return to the proper state. The complicated inter-application relationship of
cascading faults lends additional value to our environmental functions. For
instance, since rollback considers the entire system state when choosing a 
checkpoint, it recognizes that the heater should be off and windows should be open,
correcting the fault. 


\section{Limitations and Discussions}
The evaluation of the effectiveness of \sysname{}'s fault handling is limited for several reasons. First, for every run, every device used the same scheme. This means that device-based differences in optimal schemes are not shown to their full effect, as improvements in one device could be canceled out by another device that performs better under a different scheme. Second, we did not design our experiments to target the key distinguishing features of the schemes, such as adding frequent short transient faults to thwart the transient-resistant scheme or adding metrics to represent time sensitivity. Lastly, as indicated in our implementation, the randomness in fault injection, while necessary for practical evaluation, also causes fluctuation in results between device executions. This may cause the most effective scheme to change between executions. We plan to conduct more extensive experiments in a simulated IoT testbed to cover various IoT device failures through injecting more targeted faults with different types and sequences.

We restrict our evaluation to a moderate set of IoT devices and applications. It allows more tractable outputs to evaluate
our fault-handling functions and the effect of faults. More sophisticated systems with a diverse set of devices and apps, especially in the settings of industrial IoT and automobiles, would allow for a more thorough evaluation. For instance, the increase in the number of devices would lead to more cascading faults, as well as increasing the probability of device correlation and allowing some of our underutilized functions such as device redundancy to be more effective. This would also allow for suppression of entire apps to be effective, as the primary use of this technique is to stop cascading faults.


\section{Related Work}
\label{sec:relatedWork}
We present related work on recent literature that explores fault identification 
and handling in IoT systems.

\para{Fault Identification} Fault identification techniques aim to determine the presence of a fault and determine the faulty device and type. We characterize the fault identification techniques for IoT systems into three groups. First, network traffic-based techniques analyze sensor data packets to detect faults~\cite{ramanathan2005sympathy, ye2012situation}. Network-level techniques are effective at identifying fail-stop failures but ignore non-fail-stop faults. Second, redundant sensor-based techniques use data from multiple homogeneous sensors and exploit the fact that spatial sensing of close sensors should yield similar sensor states~\cite{atakli2008malicious, fang2013unifying, sharma2010sensor}. However, these systems are costly due to the requirement of multiple sensors and, more importantly, ineffective when considering that simultaneous sensor faults are common in IoT. Lastly, sensor-based techniques use data obtained from various types of sensors for fault identification. These techniques apply to environments such as smart homes where various sensors states are available and often correlated~\cite{munir2014failuresense,kodeswaran2016idea,kapitanova2012being}. While these techniques differ in scope, precision, and methodology, they limit their analysis to sensors that have binary readings and do not identify fault types. 

\para{Fault Handling} Very little work has been done in fault handling in IoT systems. The explored techniques often aim at specific environments and faults. A fault-tolerant technique~\cite{tu2018redundancy} for redundancy-free UAV sensors use imperfect replication to allow near-correct device operation in the presence of a sensor failure. While this work poses a good argument for the value of imperfect replication, the functions are specific to UAV sensors and do nothing to repair faults. Rivulet~\cite{ardekani2017rivulet} aims at removing the edge device as a single point of failure to mitigate connection faults by distributing communication to devices. The core drawback of this technique is that it does not handle the far more common sensor and actuator faults. Transactuations is a fault-handling technique introduced in~\cite{sengupta2019transactuations} to address some flaws that transactions have in IoT environments. This enhanced technique better prevents physical device states from losing synchronization with software variables. While this is useful for preserving dependencies, it does nothing to repair faults
and cannot correct errors that occur before the fault is detected. To the best of our knowledge, no current techniques develop a fault handler for diverse IoT device failures and bundles fault identification and fault handling to mitigate faults.

\section{Conclusions}
We presented \sysname{}\footnote{An extended version of this paper is available with substantially more methodology description, performance evaluation details, and commentary~\cite{IoTRepairExtended}.}, a novel fault handler for IoT deployments in the form of a flexible library that integrates with edge devices and can be customized to meet the requirements for diverse environments. \sysname{} is organized by a set of schemes to fit the handler to different deployments, and uses runtime redundant device detection and fault history to autonomously adapt to the devices and faults in the environment. With a good identification module, \sysname{} could remove more than 50.01\% or more of errors caused by faults.

\ifnum \value{technicalDoc}>0 {
\appendix{}
\section{Safety and Security Examples}
\label{apdx:SafetySec}
\begin{table*}[ht!]
\def\arraystretch{0.9}
{\small{
\resizebox{\textwidth}{!}{%
\begin{threeparttable}[b]
\begin{tabular}{|P{0.3cm}|P{1cm}|P{3.5cm}|p{10cm}|}
\hline
\multicolumn{1}{|c|}{\textbf{ID\tnote{\textdagger}}} &  \multicolumn{1}{c|}{\textbf{App ID}} & \textbf{Fault Description} & \multicolumn{1}{c|}{\textbf{Implication of Faults on IoT environment\tnote{\textdaggerdbl}}} \\ \hline \hline
\multirow{22}{*}{\circled{1}} 
& \multicolumn{1}{c|}{\multirow{3}{*}{\texttt{App1}}}& \multicolumn{1}{c|}{\multirow{3}{3.5cm}{Motion sensor fails to report its state}} & \circledempty{1} Light is off when the user is in the room \\ \cline{4-4} 
 &  &  & \circledempty{2} Light is on when the user is not in the room \\ \cline{4-4} 
 &  &  & \circledempty{3} Light is on when the user is in the room and off when not \\ \cline{2-4} 
 & \multicolumn{1}{c|}{\multirow{2}{*}{\texttt{App2}}} & \multicolumn{1}{c|}{\multirow{2}{3.5cm}{Smoke detector stuck at smoke-detected}} & \circledempty{1} Alarm sounds and door is unlocked when there is no smoke \\ \cline{4-4} 
 &  &  & \circledempty{2} and \circledempty{3}  Alarm is off and door is locked when there is no smoke \\ \cline{2-4} 
 & \multicolumn{1}{c|}{\multirow{2}{*}{\texttt{App3}}} & \multicolumn{1}{c|}{\multirow{2}{3.5cm}{Temperature sensor fails to report its state}} & \circledempty{1} Heater is on when it is warm and AC is off when it is too hot\\ \cline{4-4} 
 &  &  & \circledempty{2} and \circledempty{3} Heater is off when it is too cold and AC is off when it is too hot \\ \cline{2-4} 
 & \multicolumn{1}{c|}{\multirow{3}{*}{\texttt{App4}}} & \multicolumn{1}{c|}{\multirow{3}{3.5cm}{Leak-detector stuck at leak-not-detected}} & \circledempty{1} Alarm is off when there is a water leak \\ \cline{4-4}
 &  &  & \circledempty{2} Alarm sounds briefly when there is no leak \\ \cline{4-4} 
 &  &  & \circledempty{3} Alarm sounds when there is a leak and off when there is no water leak \\ \cline{2-4} 
 & \multicolumn{1}{c|}{\multirow{2}{*}{\texttt{App5}}} & \multicolumn{1}{c|}{\multirow{2}{3.5cm}{Door lock fails to report or change its state}} & \circledempty{1} Door is unlocked when the user is not home \\ \cline{4-4} 
 &  &  & \circledempty{2} and \circledempty{3} Door is unlocked when the user is not home  \\ \cline{2-4} 
 & \multicolumn{1}{c|}{\multirow{2}{*}{\texttt{App6}}} & \multicolumn{1}{c|}{\multirow{2}{3.5cm}{Presence sensor stuck at not-present}} & \circledempty{1} and \circledempty{2} Several SMS sent to users about insecure patio when they are home \\ \cline{4-4} 
 &  &  & \circledempty{3} No SMS sent to users about insecure patio when they are home \\ \cline{2-4} 
 & \multicolumn{1}{c|}{\multirow{2}{*}{\texttt{App7}}} & \multicolumn{1}{c|}{\multirow{2}{3.5cm}{Window opener stuck at window-open}} & \circledempty{1} and \circledempty{2} Window is open when heater or air conditioner is active \\ \cline{4-4} 
 &  &  & \circledempty{3} Window is briefly open when heater or air conditioner is active \\ \cline{2-4} 
 & \multicolumn{1}{c|}{\multirow{2}{*}{\texttt{App4}}} & \multicolumn{1}{c|}{\multirow{2}{3.5cm}{Watervalve actuator stuck at open}} & \circledempty{1} and \circledempty{2} Watervalve is open when there is a leak \\ \cline{4-4} 
 &  &  & \circledempty{3} Watervalve is briefly open when there is a leak \\ \cline{2-4} 
 & \multicolumn{1}{c|}{\multirow{2}{*}{\texttt{App9}}} & \multicolumn{1}{c|}{\multirow{2}{3.5cm}{Motion sensor stuck at motion-inactive}} & \circledempty{1}, \circledempty{2} and \circledempty{3} SMS is not sent when motion is active and user is not home \\
 &  &  & \\  \cline{2-4}
 & \multicolumn{1}{c|}{\multirow{2}{*}{\texttt{App2,10}}} & \multicolumn{1}{c|}{\multirow{2}{3.5cm}{Smoke detector stuck at smoke-detected}} & \circledempty{1} and \circledempty{2} Alarm is on, door is unlocked, and lights flash when there is no smoke\\ \cline{4-4} 
 &  &  & \circledempty{3} Alarm is off, door is locked, and lights are off when there is no smoke \\ \hline \hline
\multirow{20}{*}{\circled{2}} 
& \multicolumn{1}{c|}{\multirow{6}{*}{\texttt{App 1,5,6,8-11}}} & \multicolumn{1}{c|}{\multirow{6}{3.5cm}{Presence sensor stuck at user-home, and light switch stuck at on}} & \circledempty{1} and \circledempty{2} Door is unlocked and window is open when the user is not home (\texttt{App8}), SMS is failed to notify user when there is movement at patio (\texttt{App6}) and in the house when user is away (\texttt{App9}), light switch is on when there is no alarm (\texttt{App10}) and user is not in the room (\texttt{App1}) \\ \cline{4-4} 
&  &  & \circledempty{3} Door is unlocked when user is not home (\texttt{App8}), SMS fails to send when there is movement on the patio (\texttt{App6}) and movement in the house when user is away (\texttt{App9}), window is closed at user-specified time (\texttt{App11}) \\ \cline{2-4} 
& \multicolumn{1}{c|}{\multirow{5}{*}{\texttt{App1,3,9}}} & \multicolumn{1}{c|}{\multirow{5}{3.5cm}{Motion sensor stuck at motion-inactive, AC stuck at on, heater fails to report its state}} & \circledempty{1} and \circledempty{2} SMS is not sent when motion is active while user is not home (\texttt{App9}), light-switch off when user is in a room (\texttt{App1}), air conditioner on when temp is too low (\texttt{App3}), heater is off when temp is too low (\texttt{App3}) \\ \cline{4-4} 
&  &  & \circledempty{3} Light switch is off when user is in room(App1), AC on while temperature is too low (\texttt{App3}) \\ \cline{2-4} 
& \multicolumn{1}{c|}{\multirow{4}{*}{\texttt{App1-4,6,7,10}}} &
\multicolumn{1}{c|}{\multirow{5}{3.5cm}{Temp sensor and contact sensor fails to report its state, leak detector and light switch stuck at leak detected and on, alarm stuck at off}} & \circledempty{1} and \circledempty{2} Alarm is off when there is a leak (\texttt{App4}) and when there is smoke (\texttt{App2}), SMS is not sent to the user when there is activity at patio (\texttt{App6}), light switch is on when there is no alarm (\texttt{App10}) and when user is not in the room (\texttt{App1}), heater is off when temperature is too low and air conditioner is off when temperature is too high(App3) \\ \cline{4-4} 
&  &  & \circledempty{3} Alarm off when there is a leak (\texttt{App4}) and when there is smoke (\texttt{App2}) \\ \hline \hline 
\multirow{4}{*}{\circled{3}}  
& \multicolumn{1}{c|}{\multirow{2}{*}{\texttt{App3,7,11}}} & \multicolumn{1}{c|}{\multirow{2}{3.5cm}{Temp sensor stuck at a low temp value}} & The low temp sensor causes the heater on by \texttt{App3}, even when the real temperature is high. The heater being on then causes the window to close due to \texttt{App7}, but the window was supposed to be open due to \texttt{App11}. \\ \cline{2-4} 
& \multicolumn{1}{c|}{\multirow{2}{*}{\texttt{App1,2,10}}} & \multicolumn{1}{c|}{\multirow{2}{3.5cm}{Smoke detector stuck at smoke-detected state}} & \texttt{App2} turns on alarm when smoke is detected, which then causes \texttt{App10} to turn on the lights, but the lights should be off because of the \texttt{App1}. \\ \hline
\end{tabular}%
\begin{tablenotes}
    \item[\textdagger] \circled{1} is for single faults, \circled{2} is for multiple faults, and \circled{3} is for cascading faults. \vspace{2pt}
    \item[\textdaggerdbl] \circledempty{1} is without fault handling, \circledempty{2} uses device suppression, and \circledempty{3} uses conservative fault handling scheme.
\end{tablenotes}
\end{threeparttable}
}}}
\caption{Examples of various faults, their impact IoT environments, and results of fault handling schemes.}
\label{table:experimentResults}
\end{table*}

Table~\ref{table:experimentResults} \circled{1} presents
examples of single device faults for each IoT App. For instance, A
failure in window opener in \texttt{App7}, \texttt{Energy-Saver},
causes the window to be open incorrectly for 1,620 polls, but the
fault handler corrects the state after 300 polls, reducing incorrect
states by 81.48\%.

In Table~\ref{table:experimentResults} 
\circled{3}, we show two cascading faults that each involve three different 
applications.

\section{Power Consumptions Evaluation}
\label{apdx:PowerCons}

\begin{figure}[t!]
\centering
\includegraphics[width=\linewidth]{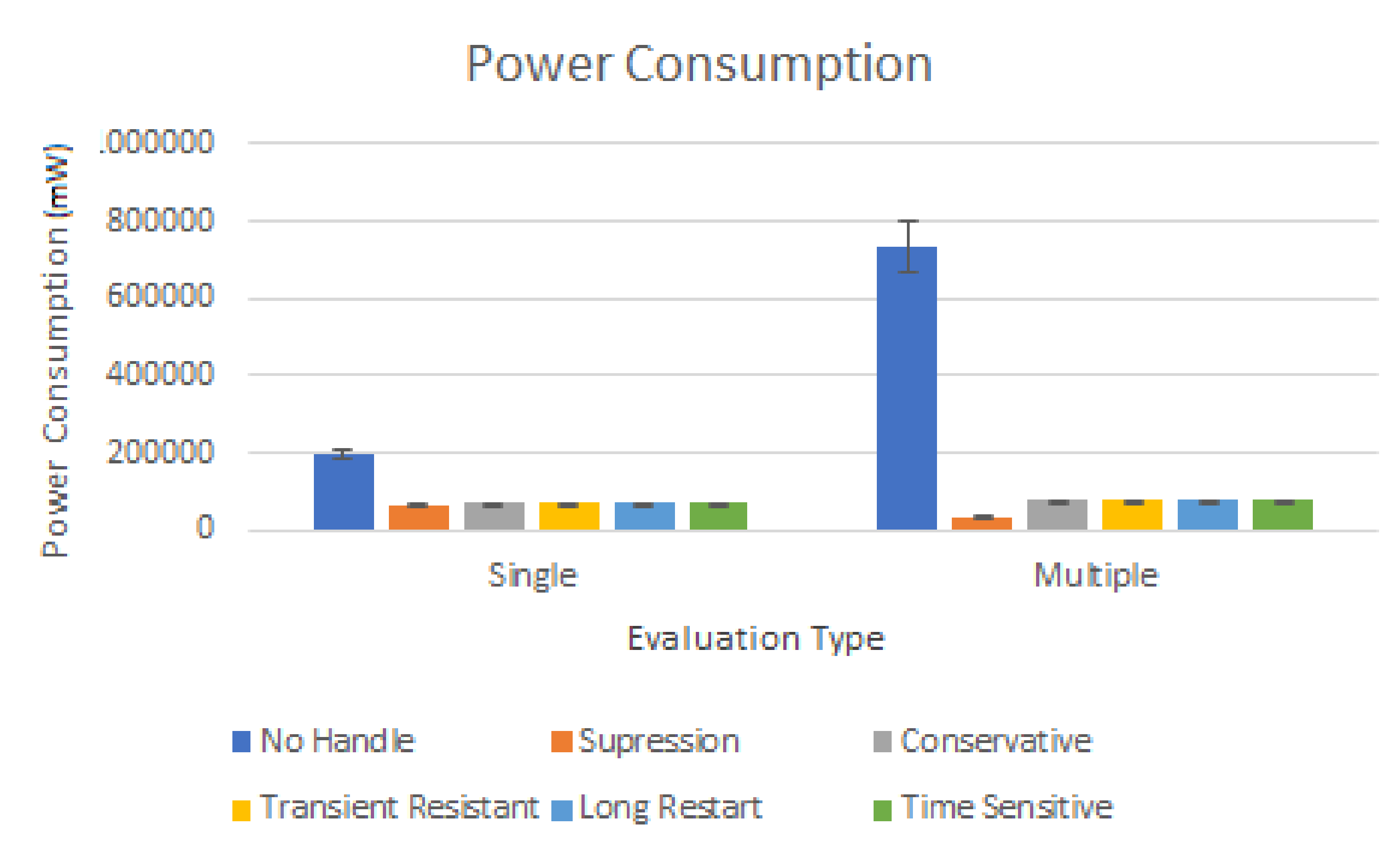}
\caption{Power overhead of Fault Handling: Consumed energy from events, actuations,
and restarts across no-handle, suppression, and schemes for our evaluations.}
\label{fig:PowerOverhead}
\end{figure} 
\label{sec:performanceOverhead}

In this set of experiments, we tracked how many events occurred, actuations were 
performed, and restarts were issued with and without fault handling. We then convert
these metrics into the power cost of each action to get the power overhead of each
execution in the evaluations. We found that power overhead for \sysname{} is
negligible. In Figure~\ref{fig:PowerOverhead}, we show the power consumption of
each experiment execution. In single fault evaluation, the worst scheme reduces
power consumption from no handler by 64.23\%. Multiple faults has a worst case of 89.36\%
decrease in 
power consumption from no handle to scheme. This decrease in power consumption is 
because faults cause many unnecessary events and actuations, and fault handling 
reduces these by suppressing and repairing the faulty devices. Relative to the power
consumption caused by unhandled faults, the additional power for restarts and 
functions like rollback are minor. Similar to the effectiveness evaluations
Overall, we find that our power overhead of our functions is small
and insignificant compared to the power consumption generated by unhandled faults.}\fi

{\footnotesize\bibliographystyle{IEEEtran}
\bibliography{refs/refs.bib}}

\end{document}